\documentclass[a4paper,11pt]{article}
\usepackage{amssymb}
\usepackage{amsmath}
\newcommand{\be}{\begin{equation}}
\newcommand{\ee}{\end{equation}}
\newcommand{\bea}{\begin{eqnarray}}
\newcommand{\eea}{\end{eqnarray}}
\newcommand{\ba}{\begin{array}}
\newcommand{\ea}{\end{array}}

\makeatletter \@addtoreset{equation}{section} \makeatother

\begin{document}

\begin{titlepage}

    \thispagestyle{empty}
    \begin{flushright}
        \hfill{SU-ITP-09/46}\\
    \end{flushright}

    \vspace{42pt}
    \begin{center}
        { \Huge{\bf On Quantum\\\vspace{7pt}Special K\"{a}hler Geometry}}

        \vspace{18pt}

        {\large{\bf Stefano Bellucci$^\clubsuit$, Alessio Marrani$^{\heartsuit}$ and \ Raju Roychowdhury$^{\spadesuit}$}}

        \vspace{15pt}

        {$\clubsuit$ \it INFN - Laboratori Nazionali di Frascati, \\
        Via Enrico Fermi 40,00044 Frascati, Italy\\
        \texttt{bellucci@lnf.infn.it}}

        \vspace{15pt}

        {$\heartsuit$ \it Stanford Institute for Theoretical Physics\\
        Department of Physics, 382 Via Pueblo Mall, Varian Lab,\\
        Stanford University, Stanford, CA 94305-4060, USA\\
        \texttt{marrani@lnf.infn.it}}

        \vspace{15pt}

        {$\spadesuit$ \it Dipartimento di Scienze Fisiche, Federico II University,\\
        Complesso Universitario di Monte S. Angelo,\\
        Via Cintia, Ed. 6, I-80126 Napoli, Italy\\
        \texttt{raju@na.infn.it}}

\end{center}

\vspace{85pt}

\begin{abstract}
We compute the effective black hole potential $V_{BH}$ of the most general $%
\mathcal{N}=2$,$d=4$ (\textit{local}) special K\"{a}hler geometry
with quantum perturbative corrections, consistent with axion-shift
Peccei-Quinn symmetry and with cubic leading order behavior.

We determine the charge configurations supporting axion-free
attractors, and explain the differences among various configurations
in relations to the presence of ``flat'' directions of $V_{BH} $ at
its critical points.

Furthermore, we elucidate the role of the sectional curvature at the
non-supersymmetric critical points of $V_{BH}$, and compute the
Riemann tensor (and related quantities), as well as the so-called
$E$-tensor. The latter expresses the non-symmetricity of the
considered quantum perturbative special K\"{a}hler geometry.

\end{abstract}

\end{titlepage}
\tableofcontents
\section{\label{Intro}Introduction}

The \textit{Attractor Mechanism} was discovered in the mid 90's \cite{FKS}-
\nocite{Strom,FK1,FK2}\cite{FGK} in the context of dynamics of scalar fields coupled to
BPS (Bogomol'ny-Prasad-Sommerfeld \cite{BPS}) black holes (BHs). In recent
years, a number of studies (see \textit{e.g.} \cite{ADFT-review}-\nocite
{Sen-review, Kallosh-review, Erice-07}\cite{BFGM2} for recent reviews, and
lists of Refs., see also \cite{Sen:2005wa}) have been devoted to the
investigation of the properties of \textit{extremal BH attractors}. This
renewed interest can be essentially be traced back to the (re)discovery of
new classes configurations of scalar fields at the BH horizon, which do not
saturate the BPS bound. When embedded into a supergravity theory, such
non-BPS configurations break all supersymmetries at the BH event horizon.

The geometry of the scalar manifold determines the various classes of BPS
and non-BPS attractors. The richest case study is provided by the theory in
which the \textit{Attractor Mechanism} was originally discovered, namely by $%
\mathcal{N}=2$, $d=4$ \textit{ungauged} supergravity coupled to $n_{V}$
Abelian vector multiplets. In such a theory, the scalar fields coordinatize
a K\"{a}hler manifold of (\textit{local}) \textit{special} type (see \textit{%
e.g.} \cite{Strominger-SKG}, \cite{CDF-review}, and \cite{Freed}, and Refs.
therein), determined by an holomorphic prepotential function $\mathcal{F}$.
In general, (\textit{local}, as understood throughout unless otherwise
noted) special K\"{a}hler (SK) geometry admits three classes of extremal BH
attractors (see \textit{e.g.} \cite{BFGM1} for an analysis in \textit{%
symmetric} SK geometry):

\begin{itemize}
\item  $\frac{1}{2}$-BPS (preserving four supersymmetries out of the eight
pertaining to asymptotical $\mathcal{N}=2$, $d=4$ superPoincar\'{e} algebra);

\item  non-BPS with non-vanising $\mathcal{N}=2$ central charge function $Z$
(shortly named non-BPS $Z\neq 0$);

\item  non-BPS with vanishing $Z$ (shortly named non-BPS $Z=0$).
\end{itemize}

\subsection{\label{Quantum-Corrs-F}Quantum Corrections to Prepotential}

Dealing with the stringy origin of $\mathcal{N}=2$, $d=4$ supergravity, the
classical prepotential $F$ receives quantum (perturbative and
non-perturbative) corrections, of polynomials or non-polynomial (usually
polylogarithmic) nature, which in some cases can spoil the holomorphicity of
$\mathcal{F}$ itself (see \textit{e.g.} \cite{Mohaupt-1}-\nocite
{CQ-N=2-BHs,Behrndt-1,Behrndt-2,Behrndt-3,Maldacena,Mohaupt-2,Cardoso-0,Mohaupt-3,R-square-1,Lust-1,Lust-2,Cardoso-1}
\cite{Cardoso-2}).

A typical (and simple) example is provided by the \textit{large volume limit}
of $CY_{3}$-compactifications of Type IIA superstrings, which determines a
SK geometry with \textit{purely cubic} $\mathcal{F}$ at the classical level.
Thus, the sub-leading nature of the quantum corrections constrains the most
general polynomial correction to $\mathcal{F}$ to be at most of degree two
in the moduli, with a priori complex coefficients. Moreover, some symmetries
can further constrain the structure of such sub-leading polynomial quantum
corrections to $\mathcal{F}$. As shown in \cite{CFG}, the only polynomial
quantum perturbative correction to classical cubic $\mathcal{F}$ which is
consistent with the perturbative (continuous) \textit{axion-shift symmetry}
\cite{Peccei-Quinn} is the constant purely imaginary term ($i=1,...,n_{V}$
throughout):
\begin{equation}
\mathcal{F}_{class}=\frac{1}{3!}d_{ijk}z^{i}z^{j}z^{k}\longrightarrow
\mathcal{F}_{quant-pert.}=\frac{1}{3!}d_{ijk}t^{i}t^{j}t^{k}+i\xi ,~\xi \in
\mathbb{R},  \label{d+i*csi}
\end{equation}
where $d_{ijk}$ is the real, constant, completely symmetric tensor defining
the cubic geometry (which is then usually named $d$-SK geometry \cite
{dWVP,dWVVP}). All other polynomial perturbative corrections (quadratic,
linear and real constant terms in the scalar fields $z^{i}$'s) can be proved
not to affect the classical $d$-SK geometry, also because the K\"{a}hler
potential is insensitive to their presence \cite{CFG}.

The explicit form of the quantum corrections to $\mathcal{F}$ depends on the
superstring theory under consideration, and non-trivial relations among the
various corrections arise due to the (perturbative and non-perturbative)
dualities relating the various superstring theories.

For instance, in a certain class of compactifications of the heterotic $%
E_{8}\times E_{8}$ superstring over $K_{3}\times T^{2}$, the whole
quantum-corrected $\mathcal{F}$ reads (see \textit{e.g.} \cite{CQ-N=2-BHs},
and Refs. therein)
\begin{eqnarray}
\mathcal{F}^{het} &=&stu-s\sum_{a=4}^{n_{V}}\left( \widetilde{t}^{a}\right)
^{2}+h_{1}\left( t,u,\widetilde{t}\right) +f_{non-pert.}\left( e^{-2\pi
s},t,u,\widetilde{t}\right) ,  \notag \\
z^{1}&\equiv& s,\;\;z^{2}\;\;\equiv\;\; t,\;\;z^{3}\;\;\equiv\;\; u,\;\;z^{a}\equiv \widetilde{t}^{a}.
\label{F-heterotic}
\end{eqnarray}
This compactifications exhibits the peculiar feature that the axio-dilaton $%
s $ belongs to a vector multiplet, and this determines the presence of ($T$%
-symmetric) quantum perturbative string-loop corrections and
non-perturbative corrections, as well. The tree-level, classical term
\begin{equation}
\mathcal{F}_{class}^{het}=stu-s\sum_{a=4}^{n_{V}}\left( \widetilde{t}%
^{a}\right) ^{2}  \label{F-het-classical}
\end{equation}
is the prepotential of the so-called \textit{generic Jordan sequence}
\begin{equation}
\frac{SU(1,1)}{U\left( 1\right) }\times \frac{SO\left( 2,n_{V}-1\right) }{%
SO\left( 2\right) \times SO\left( n_{V}-1\right) }
\end{equation}
of homogeneous \textit{symmetric} SK manifolds (see \textit{e.g.} \cite{CFG}
and \cite{BFGM1,BFGM2}, and Refs. therein). Notice that $\mathcal{F}%
_{class}^{het}$ given by Eq. (\ref{F-het-classical}) exhibits its maximal
(non-compact) symmetry, namely $SO\left( 1,n_{V}-2\right) $, pertaining to
its $d=5$ uplift. Non-renormalization theorems state that all quantum
perturbative string-loop corrections are encoded in the $1$-loop
contribution $h_{1}$, made out of a constant term, a purely cubic polynomial
term and a polylogarithmic part (see \textit{e.g.} \cite{CQ-N=2-BHs} and
Refs. therein). Finally, $f_{non-pert.}$ encodes the non-perturbative
corrections, exponentially suppressed in the limit $s\rightarrow \infty $
(see \textit{e.g.} \cite{CDLOGP1,CQ-N=2-BHs,Behrndt-3,Cardoso-2}).

As mentioned above, superstring dualities play a key role in relating the
quantum corrected $\mathcal{F}$'s in various theories. In the considered
framework, the Type IIA/heterotic duality allows for an identifications of
the relevant scalar fields (\textit{moduli} of the geometry of the internal
manifold in stringy language) such that the heterotic prepotential (\ref
{F-heterotic}) becomes structurally identical to the one determined by Type
IIA compactifications over Calabi-Yau threefolds ($CY_{3}$s). Within this
latter framework, the $\mathcal{F}$ governing the resulting low-energy $%
\mathcal{N}=2$, $d=4$ supergravity is of purely classical origin. Indeed,
there are only K\"{a}hler structure \textit{moduli}, and the axio-dilaton $s$
belongs to an hypermultiplet; this leads to no string-loop corrections, and
all corrections to the \textit{large volume limit} cubic prepotential come
from the world-sheet sigma-model \cite{Alvarez-Gaume}. In particular, as
shown in \cite{Grisaru, CDLOGP1}\textbf{,} there are no $1$-, $2$- and $3$%
-loop contributions. It is here worth pointing out that the
non-perturbative, world-sheet instanton corrections (which we will disregard
in the treatment below) spoil the continuous nature of the axion-shift
symmetry, by making it discrete \cite{Peccei-Quinn}.

Thus, the relevant part of the prepotential $\mathcal{F}$ in Type IIA
compactifications reads ($n_{V}=h_{1,1}$) \cite{CDLOGP1,HKTY,CQ-N=2-BHs}:
\begin{equation}
\mathcal{F}^{IIA}=\frac{1}{3!}\mathcal{C}_{ijk}t^{i}t^{j}t^{k}+\mathcal{W}%
_{0i}t^{i}-i\frac{\chi \zeta \left( 3\right) }{16\pi ^{3}}.  \label{F-IIA}
\end{equation}
The $\mathcal{C}_{ijk}$ are the real classical intersection numbers,
determining the classical $d$-SK geometry in the \textit{large volume limit}%
. On the other hand, the quantum perturbative contributions from $2$%
-dimensional CFT on the world-sheet are encoded only in a linear and in a
constant term:

\begin{itemize}
\item  the \textit{linear} term is determined by
\begin{equation}
\mathcal{W}_{0i}=\frac{1}{4!}c_{2}\cdot J_{i}=\frac{1}{4!}%
\int_{CY_{3}}c_{2}\wedge J_{i},
\end{equation}
which are the real expansion coefficients of the second Chern class $c_{2}$
of $CY_{3}$ with respect to the basis $J_{i}^{\ast }$ of the cohomology
group $H^{4}\left( CY_{3},\mathbb{R}\right) $, \textit{dual} to the basis of
the ($1,1$)-forms $J_{i}$ of the cohomology $H^{2}\left( CY_{3},\mathbb{R}%
\right) $. The linear term $\mathcal{W}_{0i}t^{i}$ has been shown to be
reabsorbed by a suitable symplectic transformation of the period vector;
thus, in the dual heterotic picture it has just the effect of a \textit{%
constant shift} in $Ims$ (\cite{Witten-theta}\textbf{; }see \textit{e.g.}
also discussion in \cite{CQ-N=2-BHs}).

\item  The constant term $-i\frac{\chi \zeta \left( 3\right) }{16\pi ^{3}}$($%
\equiv i\xi $ in Eq. (\ref{d+i*csi})) in (\ref{F-IIA}) is the only relevant
one, as proved in general in \cite{CFG}. It is determined by the Riemann
zeta-function $\zeta $, by the Euler character\footnote{%
For typical $CY_{3}$'s
\begin{equation*}
\left| \chi \right| \leqslant 10^{3}\Leftrightarrow \frac{\chi \zeta \left(
3\right) }{16\pi ^{3}}\sim \mathcal{O}\left( 1\right) .
\end{equation*}
This motivates the statement that attractor solutions with $\xi =0$ can be
(in certain BH charge configurations) a good approximation for the solutions
computed with $\xi \neq 0$ (see \textit{e.g.} the remark after Eq. (3.42) of
\cite{CQ-N=2-BHs}).} $\chi $ of $CY_{3}$, and it has a $4$-loop origin in
the non-linear sigma-model \cite{Alvarez-Gaume, Grisaru, CDLOGP1}. It is
worth noticing that $\chi =0$ for \textit{self-mirror} $CY_{3}$'s, such that
all have $\xi =0$. Furthermore, some arguments lead to argue that (%
\textit{at least}) for some particular \textit{self-mirror} models (such as
the so-called FHSV one \cite{FHSV} and the octonionic magic \cite
{Ferrara-Bianchi}), non-perturbative, world-sheet instanton corrections
vanish, as well (see e.g. discussion in Sects. 12 and 13 of \cite
{Gunaydin-lectures}, and Refs. therein). As a consequence, such models, up
to suitable symplectic transformations of the period vector, would have
their classical \textit{cubic} prepotential unaffected by any perturbative
and non-perturbative correction.
\end{itemize}

It should be here pointed out that $CY_{3}$-compactifications of Type
IIB do not admit a large volume limit; moreover, in Type IIB the \textit{%
Attractor Eqs.} only depend on the complex structure \textit{moduli} (which
in supergravity description are the scalars of the $\mathcal{N}=2$ vector
multiplets). The solutions to $\mathcal{N}=2$, $d=4$ \textit{Attractor Eqs.}
for the resulting SK geometries were studied (in proximity of the
Landau-Ginzburg point) in \cite{BFMY} for the particular class of Fermat $%
CY_{3}$'s with $n_{V}=1$, and in \cite{Misra1} for a particular $CY_{3}$
with $n_{V}=2$.\medskip

In \cite{BFMS1}, extending the BPS analysis of \cite{CQ-N=2-BHs}, the $%
\mathcal{N}=2$, $d=4$ \textit{Attractor Eqs.} were studied in the simplest
case of perturbative quantum corrected $d$-SK geometry, namely in the SK
geometry with $n_{V}=1$ scalar fields, described (in a \textit{special
coordinates}) by the holomorphic K\"{a}hler gauge-invariant prepotential%
\footnote{%
In \cite{BFMS1} $\xi $ was named $\lambda $.}
\begin{equation}
\mathcal{F}=t^{3}+i\xi ,  \label{t^3+i*csi}
\end{equation}
which, up to overall rescaling, is nothing but $\mathcal{F}_{quant-pert.}$
of Eq. (\ref{d+i*csi}) for $n_{V}=1$. Despite the (apparently) minor
correction to the classical prepotential, in \cite{BFMS1} new phenomena,
absent in the classical limit $\xi =0$, were observed:

\begin{itemize}
\item  The \textit{``separation''} of attractors, namely the existence of
multiple stable solutions to the Attractor Eqs. (for a given BH charge
configuration). This can be ultimately related to the existence of \textit{%
basins of attractions }\cite{Kal1,Kal2,Zhukov}, specified by suitable
\textit{``area codes''} \cite{Moore} (\textit{i.e.} asymptotical boundary
conditions) in the radial evolution dynamics of scalar fields in the
extremal BH background.

\item  The \textit{``transmutation''} of attractors, namely the change in
the supersymmetry preserving features of stable critical points of $V_{BH}$,
depending on the value of the quantum parameter $\xi $, suitably \textit{%
``renormalized''} in terms of the relevant BH charges. For example, by
varying such a \textit{``renormalized''} quantum parameter, a $\frac{1}{2}$%
-BPS attractor becomes non-BPS (and vice versa). This can ultimately be
related to the lack of an orbit structure in the space of Bh charges; this
is no surprise, by noticing that the SK geometry determined by $\mathcal{F}$
given by Eq. (\ref{t^3+i*csi}) is generally not symmetric nor homogeneous.
\end{itemize}

\subsection{\label{Flat-Dirs-V}Critical \textit{``Flat'' }Directions of
Black Hole Potential}

Let us now shortly recall the fundamentals of the \textit{Attractor Mechanism%
}. In the critical implementation given in \cite{FGK}, the \textit{Attractor
Mechanism related }extremal BH attractors to \textit{stable} critical points%
\footnote{%
The subscript ``$H$'' denotes the value at the event horizon of the
considered extremal BH.} $z_{H}^{i}$ of a suitably defined BH effective
potential $V_{BH}$:
\begin{equation}
z_{H}\left( Q\right) :\left. \frac{\partial V_{BH}\left( z,Q\right) }{%
\partial z^{i}}\right| _{z=z_{H}\left( Q\right) }=0  \label{Mon-1}
\end{equation}
where $Q$ denotes the $Sp\left( 2n_{V}+2,\mathbb{R}\right) $-vector of
magnetic and electric BH charges (see Eq. (\ref{Q}) below). The $n_{V}$
complex Eqs. (\ref{Mon-1}) are usually called \textit{Attractor Equations}.
Then, a critical point $z_{H}\left( Q\right) $ is an \textit{attractor} in
strict sense \textit{iff} the (Hermitian) $2n_{V}\times 2n_{V}$ Hessian
matrix $\mathcal{H}^{V_{BH}}$ of $V_{BH}$ evaluated at $z_{H}\left( Q\right)
$ is \textit{positive definite}:
\begin{equation}
\left. \mathcal{H}^{V_{BH}}\right| _{z=z_{H}\left( Q\right) }\geqslant 0,
\label{Mon-2}
\end{equation}
with ``$\geqslant 0$'' here expressing the non-negativity of the $2n_{V}$
eigenvalues.\smallskip

As shown in \cite{Ferrara-Marrani-1,ferrara4}, in $\mathcal{N}=2$, $d=4$
\textit{ungauged} supergravities with homogeneous (not necessarily
symmetric) SK manifolds (as well as in $\mathcal{N}>2$-extended \textit{%
ungauged} $d=4$ supergravities, which we however do not consider here) the
critical matrix $\left. \mathcal{H}^{V_{BH}}\right| _{\partial V_{BH}=0}$
has the following general signature: all strictly positive eigenvalues, up
to some eventual vanishing eigenvalues (\textit{massless Hessian modes}),
which have been proved to be \textit{``flat''} directions of $V_{BH}$ itself.

Thus, one can claim that in all homogeneous SK geometries the critical
points of $V_{BH}$ satisfying the ``non-degeneracy'' condition
\begin{equation}
\left. V_{BH}\right| _{\partial V_{BH}=0}\neq 0  \label{non-deg-cond}
\end{equation}
are all \textit{stable}, up to some eventual \textit{``flat''} directions.
Such directions of the SK scalar manifold $\mathcal{M}_{SK}$ coordinatize
the so-called \textit{moduli space} $\frak{M}\subsetneq \mathcal{M}$ of the
considered (class of) solution(s) to Eqs. (\ref{Mon-1}). In other words,
such \textit{``flat''} directions span a subset of the scalar fields which
is \textit{not} stabilized by the Attractor Eqs. (\ref{Mon-1}) at the BH
event horizon in terms of the BH charges $Q$. It is worth pointing out that,
somewhat surprisingly, the existence of \textit{``flat''} directions at the
critical points of $V_{BH}$ does \textit{not} plague the thermodynamical
macroscopic description of \textit{extremal} BHs with inconsistencies.
Indeed, at the considered class of critical points, $V_{BH}$ does
\textit{not} actually turn out to depend on the unstabilized scalars;
therefore, through the relation \cite{FGK}
\begin{equation}
S_{BH}\left( Q\right) =\pi \left. V_{BH}\right| _{\partial V_{BH}=0},
\label{Mon-3}
\end{equation}
the BH entropy $S_{BH}$ can be consistently defined. Notice that the
condition (\ref{non-deg-cond}) implies the (classical) \textit{Attractor
Mechanism}\footnote{\textit{Attractor Mechanism} can be consistently
implemented at the quantum level, \textit{at least} in some frameworks, for
instance within the so-called entropy function formalism (see \textit{e.g.}
\cite{Sen-review} and Refs. therein, see also \cite{Sen:2005wa}). See also
\cite{BFSY-1} (and Refs. therein) for recent developments concerning \textit{%
Attractor Mechanism} and higher derivatives corrections to Einstein
(super)gravity theories.}\textit{\ }to work only for the so-called \textit{%
``large''} BHS, \textit{i.e.} for those BHs with non-vanishing classical
entropy.

As known since \cite{FGK}, \textit{``flat''} directions cannot arise at $%
\frac{1}{2}$\_BPS critical points of $V_{BH}$. This is no more true for the
remaining two classes of non-supersymmetric critical points, namley for
non-BPS $Z\neq 0$ and non-BPS $Z=0$ ones \cite{Ferrara-Marrani-1,ferrara4}.
Tables 2 and 3 of \cite{ferrara4} respectively list the \textit{moduli
spaces }of non-BPS $Z\neq 0$ and non-BPS $Z=0$ attractors for symmetric SK
geometries, whose classification is known after \cite{CVP} (see also \cite
{dWVVP} and \cite{CFG}, as well as Refs. therein). Let us mention that
non-BPS $Z\neq 0$ \textit{moduli spaces} are nothing but the symmetric
\textit{real special} scalar manifolds of the corresponding $\mathcal{N}=2$,
$d=5$ supergravity.

\subsection{\label{Removal/Survival}Quantum Removal/Survival of Critical
\textit{``Flat''} Directions}

It should be pointed out clearly that the issue of the \textit{``flat''}
directions of $V_{BH}$ at its critical points, reported in Subsect. \ref
{Flat-Dirs-V} hold only at the classical, Einstein supergravity level. It is
conceivable that such \textit{``flat''} directions are removed by quantum
(perturbative \textit{and/or} non-perturbative) corrections. Consequently,
at the quantum (perturbative and/or non-perturbative) regime\textit{, no
moduli spaces for attractor solutions might exist at all} (and also the
actual attractive nature of the critical points of $V_{BH}$\ might be
destroyed). \textit{However, this might not be the case for }$\mathcal{N}=8$%
, or for some particular charge configurations in $\mathcal{N}<8$
supergravities (see below).

By relating the issues reported in Subsects. \ref{Quantum-Corrs-F} and \ref
{Flat-Dirs-V}, one might thus ask about the fate of classical \textit{%
``flat''} directions of $V_{BH}$ at its (non-BPS) critical points, in
presence of quantum (perturbative \textit{and/or} non-perturbative)
corrections to the prepotential $\mathcal{F}$ of SK geometry.

This issue, crucial in order to understand the features of the \textit{%
Attractor Mechanism} in the quantum regime (and thus its consistent
embedding in the high-energy theories whose supergravity is an effective
low-energy limit, namely superstrings and $M$-theory), was started to be
investigated in \cite{BFMS2}, and it is the object of the investigation
carried out in the present paper.

Let us start by recalling the simplest symmetric $d$-SK geometries, and
their eventual non-BPS\ \textit{``flat''} directions. For our purpose, it
will suffice to consider only the so-called $t^{3}$ and $st^{2}$ models:

\begin{itemize}
\item  The $t^{3}$ model is based on the rank-$1$ symmetric SK manifold
\begin{equation}
\frac{SU\left( 1,1\right) }{U\left( 1\right) },
\end{equation}
endowed with prepotential ($z^{1}\equiv t$, $Imt<0$)
\begin{equation}
\mathcal{F}=t^{3},
\end{equation}
which is the classical limit $\xi \rightarrow 0$ of Eq. (\ref{t^3+i*csi}).
As yielded by the analysis of \cite{CVP}, it is an isolated case in the
classification of symmetric SK geometries (see also \cite{CFG}).
Furthermore, such a model can also be conceived as the \textit{``}$\mathit{%
s=t=u}$\textit{\ degeneration'' }of the so-called $stu$ model \cite{Duff-stu}%
-\nocite{BKRSW,Shmakova,TT1,Saraikin-Vafa-1,TT2,BMOS-1}\cite{stu-unveiled},
or equivalently as the \textit{``}$\mathit{s=t}$\textit{\ degeneration'' }of
the so-called $st^{2}$ model (see below). Beside the $\frac{1}{2}$-BPS
attractors, the $t^{3}$ model (whose $d=5$ uplift is \textit{pure} $\mathcal{%
N}=2$, $d=5$ supergravity) admits only non-BPS $Z\neq 0$ critical points of $%
V_{BH}$ with no \textit{``flat''} directions (and thus no associated \textit{%
moduli space}) \cite{BFGM1,ferrara4}.

\item  The $st^{2}$ model is based on the rank-$2$ symmetric SK manifold
\begin{equation}
\left( \frac{SU\left( 1,1\right) }{U\left( 1\right) }\right) ^{2},
\end{equation}
endowed with prepotential ($z^{1}\equiv s$, $z^{2}\equiv t$, $Ims<0$, $Imt<0$%
)
\begin{equation}
\mathcal{F}=st^{2}.
\end{equation}
It has one non-BPS $Z\neq 0$ \textit{``flat''} direction, spanning the
\textit{moduli space} $SO\left( 1,1\right) $ (namely, the scalar manifold of
the $st^{2}$ model in $d=5$), but \textit{no} non-BPS $Z=0$ \textit{``flat''}
directions at all. Such a model is the smallest (\textit{i.e.} the
fewest-moduli) \textit{symmetric} model exhibiting a non-BPS $Z\neq 0$
\textit{``flat''} direction. Remarkably, the $st^{2}$ model constitutes the
unique example of homogeneous $d$-SK geometry with $n_{V}=2$ scalar fields
\cite{dWVVP,dWVP}. Furthermore, as evident from the structure of the cubic
norm (see \textit{e.g.} the discussion in \cite{dWVVP}, as well as Eq.
(3.2.3) and Sect. 5 of \cite{AFMT1}), the $st^{2}$ model is the unique $%
n_{V}=2$ SK geometry to be uplifted to anomaly-free \textit{pure} $\left(
1,0\right) $, $d=6$ supergravity (at least in presence of neutral matter).
\end{itemize}

As mentioned at the end of Subsect. \ref{Quantum-Corrs-F}, the
non-homogeneous model ``$t^{3}+i\xi $'' (with prepotential given by Eq. (\ref
{t^3+i*csi})) was studied in \cite{BFMS1}. The ``$t^{3}+i\xi $'' model can
be conceived as the prototype of quantum perturbative corrected SK geometry,
because it is the $n_{V}=1$ SK geometry with the most general quantum
perturbative correction consistent with the (continuous, perturbative)
axion-shift symmetry \cite{CFG}. However, since the $t^{3}$ model has no
non-BPS \textit{``flat''} directions at all, the study performed in \cite
{BFMS1} is not relevant for the aforementioned issue of the fate of the
\textit{moduli spaces} of classical attractors in the quantum regime.

From the above analysis, the $st^{2}$ model is the simplest example of SK
geometry in which the study of the fate of classical non-BPS $Z\neq 0$
moduli space can be investigated in quantum perturbative regime, namely
considering the ``$st^{2}+i\xi $'' model, whose prepotential in special
coordinates reads
\begin{equation}
\mathcal{F}=st^{2}+i\xi .  \label{st^2+i*csi}
\end{equation}
Notice that Eq. (\ref{st^2+i*csi}) is the unique homogeneous $n_{V}=2$
determination of Eq. (\ref{d+i*csi}). Such a study was performed in \cite
{BFMS2}, within the (supergravity analogues of the) so-called \textit{%
magnetic} ($D0-D4$), \textit{electric} ($D2-D6$) and $D0-D6$ BH charge
configurations. As somewhat intuitively expected, in the magnetic and
electric configurations the classical non-BPS $Z\neq 0$ moduli space $%
SO(1,1) $ was shown not to survive after the introduction of the quantum parameter
$\xi \neq 0$. Interestingly, the investigation of \cite{BFMS2} showed the
that the quantum removal of classical \textit{``flat''} directions occurs
more often towards \textit{repeller} directions (thus destabilizing the
whole critical solution, and \textit{destroying the attractor in strict sense%
}), rather than towards \textit{attractive} directions.

Surprisingly, the study of \cite{BFMS2} also revealed that the $D0-D6$
configuration exhibits a qualitatively different phenomenon, namely that the
non-BPS $Z\neq 0$ classical \textit{``flat''} direction \textit{survives}
the considered quantum perturbative corrections effectively encoded in the ``%
$+i\xi $'' term in Eq.(\ref{st^2+i*csi}), despite acquiring a \textit{%
non-vanishing} axionic part.

\subsection*{Aim and Plan of Paper}

This unexpected fact was not completely understood in \cite{BFMS2}, and it
is the starting point of the present investigation, which aims at thoroughly
investigating, within the effective BH potential formalism, the $d$-SK
geometries with the most general quantum perturbative correction consistent
with continuous Peccei-Quinn axion-shift symmetry, namely the SK geometries
with prepotential (in \textit{special coordinates}) given by Eq. (\ref
{d+i*csi}). As already found in the simple cases investigated in \cite{BFMS1}
($n_{V}=1$) and \cite{BFMS2} ($n_{V}=2$), the Attractor Eqs. (especially the
non-supersymmetric ones) cannot be solved analytically for a generic BH
charge configuration, because they turn out to be algebraic Eqs. of high ($%
>4 $) order. However, by explicitly computing $V_{BH}$ for the prepotential (%
\ref{d+i*csi}), we will explain the peculiarity of the $D0-D6$ configuration
as being , in presence of $\xi \neq 0$, somewhat the \textit{``minimal''}
configuration which does not support \textit{axion-free} attractor
solutions. In light of new results concerning the relation between the
so-called \textit{sectional curvature of matter charges} at the
(non-supersymmetric) critical points of $V_{BH}$ and the BH entropy $S_{BH}$%
, we will then compute the relevant tensors characterizing the quantum SK
geometry (\ref{d+i*csi}), namely the Riemann tensor and related
contractions, and the $E$-tensor.\smallskip

The plan of the paper is as follows.\smallskip

In Sect. \ref{V_BH} we explicitly compute the effective BH potential $V_{BH}$
for the most general quantum perturbatively corrected SK geometry consistent
with continuous axion-shift symmetry, namely the one with prepotential (\ref
{d+i*csi}), in full generality, \textit{i.e.} for an arbitrary number $n_{V}$
of vector multiplets and for a generic configuration $Q$ of BH charges. We
then determine the \textit{axion-free}-supporting Bh\ charge configurations,
commenting on the role of $D0-D6$, and (partially) explaining the findings
of \cite{BFMS2}.

Sect. \ref{Z-and-DZ} is devoted to the computation of the $\mathcal{N}=2$
central charge $Z$ and the related \textit{matter charges }$D_{i}Z$ in the
considered framework. Such a computations allows one to draw some general
statements on the $\frac{1}{2}$-BPS solutions, connecting to the few results
already known from literature \cite{CQ-N=2-BHs}.

In Sect. \ref{E-Tensor} the role of the so-called $E$-tensor in SK geometry
(and in the \textit{Attractor Mechanism} within) is recalled, and its
explicit computations for the geometry (\ref{d+i*csi}) is presented. By
performing the classical limit $\xi \rightarrow 0$, the $E$-tensor for a
generic $d$-SK geometry is explicitly obtained. The factorizability of some
functional dependences for the classical $E$-tensor is explicitly found,
highlighting the possibility to uplift the theory to $d=5$. The same does
not happen when $\xi \neq 0$, thus confirming the well known fact that only $%
d$-SK geometry admits an uplift to $d=5$ (see \textit{e.g.} \cite{CFM1} and
Refs. therein).

Then, in Sect. \ref{Sect-Curv} a number of original results are derived,
pointing out the role of the so-called \textit{sectional curvature of matter
charges} $\mathcal{R}$ in the theory of non-supersymmetric attractors.
Indeed, $\mathcal{R}$ vanishes at $\frac{1}{2}$-BPS attractors, but it is
proportional to the critical value of $V_{BH}$ (and thus, through Eq. (\ref
{Mon-3}), to the BH entropy $S_{BH}$). In particular, in symmetric SK
geometries it has the same sign of the quartic invariant $\mathcal{I}_{4}$
at non-BPS $Z\neq 0$ critical points, whereas it is opposite to (the double
of) $\mathcal{I}_{4}$ at non-BPS $Z=0$ critical points, thus being
strictly negative in both cases.

Since $\mathcal{R}$ is nothing but the contraction of the Riemann tensor
with the matter charges vectors (\textit{i.e.} with covariant derivatives of
$Z$ itself), interesting role of $\mathcal{R}$ at non-supersymmetric
critical points of $V_{BH}$ elucidated in Sect. \ref{Sect-Curv} calls for an
explicit computation of the Riemann tensor itself. This is carried out in
Sect. \ref{Riemann-Tensor}, where also the Ricci tensor and the Ricci scalar
curvature are determined. We proceed by exploiting two different approaches,
one merely based on K\"{a}hler geometry (Subsect. \ref{First-Approach}) and
the second one (Subsect. \ref{Second-Approach}) based instead on the
fundamental constraints of SK geometry (see Eq. (\ref{SKG-constraints}
below). We explicitly show the equivalence of these two approaches, by
shortly commenting on the results of \cite{CVP} and on the eventual
(unlikely) Einstein nature of the SK geometries (\ref{d+i*csi}).

Finally, Sect. \ref{Conclusion} makes a brief comment and outlook, and lists
some of the various open issues, originated or highlighted \ by the present
investigation, which we leave for future study.

Three Appendices conclude the paper. They respectively contain computational
details concerning $V_{BH}$ (App. A), the $E$-tensor (App. B), and the
Riemann tensor (App. C).

\section{\label{V_BH}Effective Black Hole Potential}

As recalled in previous Section and as firstly found in \cite{CFG}, the most
general holomorphic prepotential with leading \textit{cubic} behavior
consistent with (perturbative, continuous) Peccei-Quinn axion-shift symmetry
\cite{Peccei-Quinn}, and which affects the K\"{a}hler potential $K$ of SK
geometry, reads
\begin{equation}
F(X;\xi )=\frac{1}{3!}d_{ijk}\frac{X^{i}X^{j}X^{k}}{X^{0}}+i\xi (X^{0})^{2},
\label{prepot}
\end{equation}
which is nothing but Eq. (\ref{d+i*csi}) before projectivizing, and before
switching to \textit{special coordinates} and suitably fixing the K\"{a}hler
gauge (see below). Let us recall once again that $i=1,...,n_{V}$ throughout (%
$n_{V}$ denoting the number of Abelian vector multiplets coupled to the $%
\mathcal{N}=2$, $d=4$ supergravity one), and $\xi \in \mathbb{R}$.

Aim of the present Section is to compute the effective BH potential $V_{BH}$
for the SK geometry determined by the holomorphic prepotential (\ref{prepot}%
). Below we will present only the main formul\ae , addressing the reader to
Appendix A for the further details of the calculations.\smallskip\

A general formula determining the kinetic vector matrix $\mathcal{N}%
_{\Lambda \Sigma }$ reads (see \textit{e.g.} \cite{FerBig})\textbf{\ }($%
\Lambda =0,1,...,n_{V}$ throughout)
\begin{equation}
\mathcal{N}_{\Lambda \Sigma }=\overline{\mathcal{F}}_{\Lambda \Sigma }+2i%
\frac{Im\left( \mathcal{F}_{\Lambda \Omega }\right) Im\left( \mathcal{F}%
_{\Sigma \Delta }\right) X^{\Omega }X^{\Delta }}{Im\left( \mathcal{F}%
_{\Theta \Xi }\right) X^{\Theta }X^{\Xi }}.  \label{N-matrix}
\end{equation}
After projectivizing, it is convenient to switch to the so-called \textit{%
special} coordinates (see \textit{e.g.} \cite{CDF-review} and Refs.
therein), defined by ($a=1,...,n_{V}$)
\begin{equation}
e_{i}^{a}\left( z\right) \equiv \frac{\partial \left( \frac{X^{a}}{X^{0}}%
\right) }{\partial z^{i}}\equiv \delta _{i}^{a},  \label{special-coords}
\end{equation}
where ($x^{i}$, $\lambda ^{i}\in \mathbb{R}$)
\begin{equation}
z^{i}\equiv x^{i}-i\lambda ^{i}  \label{z-split-def}
\end{equation}
are the $n_{V}$ complex scalar fields, and further suitably fix the
K\"{a}hler gauge as
\begin{equation}
X^{0}\equiv 1.  \label{Kgf}
\end{equation}
Within such a framework, one can thus write:
\begin{equation}
Im\left[ \mathcal{F}_{\Lambda \Sigma }(z;\xi )\right] =\left(
\begin{array}{ccc}
\frac{1}{3}d_{ijk}Im\left( z^{i}z^{j}z^{k}\right) +2\xi &  & -\frac{1}{2}%
d_{jkl}Im\left( z^{k}z^{l}\right) \\
&  &  \\
-\frac{1}{2}d_{ikl}Im\left( z^{k}z^{l}\right) &  & d_{ijk}Im\left(
z^{k}\right)
\end{array}
\right) ,  \label{imfsymplectic}
\end{equation}
and the block components of $\mathcal{N}_{\Lambda \Sigma }$ are computed in
Appendix A, the final results being given by Eqs. (\ref{N00}), (\ref{N0i})
and (\ref{Nij}).

In order to compute the $V_{BH}$ governing the \textit{Attractor Mechanism}
\cite{FKS,Strom,FK1,FK2,FGK}, it is worth recalling that in $\mathcal{N}=2$,
$d=4$ \textit{ungauged} Maxwell-Einstein supergravity the following
expression holds \cite{FK1,FK2,CDF-review}:
\begin{equation}
V_{BH}=\left| Z\right| ^{2}+g^{j\overline{j}}\left( D_{j}Z\right) \overline{D%
}_{\overline{j}}\overline{Z},  \label{VBH1}
\end{equation}
where $Z$ is the $\mathcal{N}=2$ \textit{central charge} function. On the
other hand, an equivalent (and independent on the number of supercharge
generators) expression of $V_{BH}$ reads \cite{FGK}
\begin{equation}
V_{BH}=-\frac{1}{2}Q^{T}\mathcal{M}\left( \mathcal{N}\right) Q.  \label{VBH2}
\end{equation}
$Q$ is the $\left( Sp\left( 2n_{V}+2,\mathbb{R}\right) \right) $-vector of
magnetic and electric charges, which in the \textit{special} \textit{%
coordinate} basis of $\mathcal{N}=2$ theory reads as follows:
\begin{equation}
Q=\left(
\begin{array}{c}
p^{0} \\
p^{i} \\
q_{0} \\
q_{i}
\end{array}
\right) .  \label{Q}
\end{equation}
The $\left( 2n_{V}+2\right) \times \left( 2n_{V}+2\right) $ real symmetric
symplectic matrix $\mathcal{M}\left( \mathcal{N}\right) $ is defined as \cite
{CDF-review,FK1,FK2}
\begin{eqnarray}
\mathcal{M}\left( \mathcal{N}\right) &=&\mathcal{M}\left( Re\left( \mathcal{N%
}\right) ,Im\left( \mathcal{N}\right) \right) \equiv  \notag \\
&&  \notag \\
&\equiv &\left(
\begin{array}{ccc}
Im\left( \mathcal{N}\right) +Re\left( \mathcal{N}\right) \left( Im\left(
\mathcal{N}\right) \right) ^{-1}Re\left( \mathcal{N}\right) &  & -Re\left(
\mathcal{N}\right) \left( Im\left( \mathcal{N}\right) \right) ^{-1} \\
&  &  \\
-\left( Im\left( \mathcal{N}\right) \right) ^{-1}Re\left( \mathcal{N}\right)
&  & \left( Im\left( \mathcal{N}\right) \right) ^{-1}
\end{array}
\right) .  \notag \\
&&
\end{eqnarray}

Thus, in order to compute $V_{BH}$ for the $\mathcal{N}=2$, $d=4$ specified
by the (perturbative) quantum corrected holomorphic prepotential (\ref
{prepot}), one has to compute the inverse of matrix $Im\mathcal{N}_{\Lambda
\Sigma }$.

It is also convenient to further simplify the notation, by recalling the
definitions used in \cite{CFM1}, and suitably changing them\footnote{%
Notice that in \cite{CFM1} a different notation was used, namely:
\begin{eqnarray*}
\kappa _{ij} &\equiv &d_{ijk}\lambda ^{k}; \\
\kappa _{i} &\equiv &d_{ijk}\lambda ^{j}\lambda ^{k}; \\
\kappa &\equiv &d_{ijk}\lambda ^{i}\lambda ^{j}\lambda ^{k}=6\nu ; \\
\kappa ^{ij}\kappa _{jl} &\equiv &\delta _{l}^{i}.
\end{eqnarray*}
} (taking into account the presence of effective quantum parameter $\xi $):
\begin{eqnarray}
d_{ij} &\equiv &d_{ijk}\lambda ^{k}; \\
d_{i} &\equiv &d_{ijk}\lambda ^{j}\lambda ^{k}; \\
\nu &\equiv &\frac{1}{3!}d_{ijk}\lambda ^{i}\lambda ^{j}\lambda ^{k};
\label{nu} \\
\widetilde{\nu } &\equiv &\nu +\frac{1}{4}\xi ; \\
h_{ij} &\equiv &d_{ijk}x^{k}; \\
h_{i} &\equiv &d_{ijk}x^{j}x^{k}; \\
h &\equiv &d_{ijk}x^{i}x^{j}x^{k},
\end{eqnarray}
thus \textit{e.g.} yielding
\begin{equation}
h_{ij}\lambda ^{i}\lambda ^{j}=d_{i}x^{i}.
\end{equation}
By further introducing \textit{``rescaled dilatons''} \cite{CFM1}
\begin{equation}
\widehat{\lambda }^{i}\equiv \frac{\lambda ^{i}}{\nu ^{1/3}}\Rightarrow
\frac{1}{3!}d_{ijk}\widehat{\lambda }^{i}\widehat{\lambda }^{j}\widehat{%
\lambda }^{k}=1,  \label{def-resc-dils}
\end{equation}
one can then define the following quantities:
\begin{eqnarray}
\widehat{d}_{ij} &\equiv &d_{ijk}\widehat{\lambda }^{k}=\nu ^{-1/3}d_{ij};
\label{deff-1} \\
\widehat{d}_{i} &\equiv &d_{ijk}\widehat{\lambda }^{j}\widehat{\lambda }%
^{k}=\nu ^{-2/3}d_{i}.  \label{deff-2}
\end{eqnarray}

Let us also recall Eq. (31) of \cite{BFMS1}, giving the expression of
covariant metric tensor $g_{i\overline{j}}$ for the prepotential (\ref
{prepot}) within the assumptions (\ref{special-coords})-(\ref{Kgf}):
\begin{equation}
g_{i\overline{j}}=g_{ij}=-\frac{1}{4(\nu -\frac{1}{2}\xi )}\left[ d_{ij}-%
\frac{d_{i}d_{j}}{4(\nu -\frac{1}{2}\xi )}\right] =-\frac{\nu ^{1/3}}{4(\nu -%
\frac{1}{2}\xi )}\left[ \widehat{d}_{ij}-\frac{\nu \widehat{d}_{i}\widehat{d}%
_{j}}{4(\nu -\frac{1}{2}\xi )}\right] .  \label{covmetric}
\end{equation}
The corresponding inverse metric ($g^{ij}g_{jk}\equiv \delta _{k}^{i}$) is
computed as
\begin{equation}
g^{i\overline{j}}=g^{ij}=-4(\nu -\frac{1}{2}\xi )\left[ d^{ij}-\frac{\lambda
^{i}\lambda ^{j}}{2(\nu +\xi )}\right] =-4(\nu -\frac{1}{2}\xi )\left[ \nu
^{-1/3}\widehat{d}^{ij}-\frac{\nu ^{2/3}\widehat{\lambda }^{i}\widehat{%
\lambda }^{j}}{2(\nu +\xi )}\right]  \label{contrmetric}
\end{equation}
where
\begin{equation}
d^{ij}d_{jk}\equiv \delta _{k}^{i}\Leftrightarrow \widehat{d}^{ij}\widehat{d}%
_{jk}\equiv \delta _{k}^{i}.
\end{equation}
The limit $\xi \rightarrow 0$ consistently yields the analogue results for $%
d $-SKG, given by Eqs. (2.4) and (2.6) of \cite{CFM1}:
\begin{eqnarray}
\lim_{\xi \rightarrow 0}g_{ij} &=&-\frac{1}{4}\nu ^{-2/3}\left( \widehat{d}%
_{ij}-\frac{\widehat{d}_{i}\widehat{d}_{j}}{4}\right) \equiv \breve{g}_{ij};
\label{3} \\
\lim_{\xi \rightarrow 0}g^{ij} &=&2\nu ^{2/3}\left( \widehat{\lambda }^{i}%
\widehat{\lambda }^{j}-2\widehat{d}^{ij}\right) \equiv \breve{g}^{ij},~~%
\breve{g}^{ij}\breve{g}_{jk}\equiv \delta _{k}^{i},  \label{4}
\end{eqnarray}
where $\breve{g}_{ij}$ and $\breve{g}^{ij}$ denote the covariant and
contravariant \textit{classical} ($\xi \rightarrow 0$) metric tensor.

After long but straightforward computations (detailed in Appendix A), the
following explicit expression of $V_{BH}$ is achieved:
\begin{eqnarray}
&&
\begin{array}{l}
V_{BH}\left( x^{j},\widehat{\lambda }^{j},\nu ;Q,\xi \right) =\frac{1}{2%
\tilde{\nu}}\left( 1-\left( \frac{3}{4}\right) ^{2}\frac{\xi ^{2}}{\tilde{\nu%
}^{2}}\right) ^{-1}\cdot \\
\\
\cdot \left[
\begin{array}{l}
\left[
\begin{array}{l}
\tilde{\nu}^{2}\left( 1-\left( \frac{3}{4}\right) ^{2}\frac{\xi ^{2}}{\tilde{%
\nu}^{2}}\right) ^{2}+\frac{h^{2}}{36}-\frac{1}{8}\xi \frac{\nu ^{2/3}}{%
\tilde{\nu}}h\widehat{d}_{i}x^{i}+ \\
\\
+\left( \frac{3}{8}\right) ^{2}\xi ^{2}\frac{\nu ^{4/3}}{\tilde{\nu}^{2}}%
\left( \widehat{d}_{i}x^{i}\right) ^{2}-\tilde{\nu}^{2}\left( 1-\left( \frac{%
3}{4}\right) ^{2}\frac{\xi ^{2}}{\tilde{\nu}^{2}}\right) 12\mathcal{A}+ \\
\\
-\frac{1}{12}\left( 1-\left( \frac{3}{4}\right) ^{2}\frac{\xi ^{2}}{\tilde{%
\nu}^{2}}\right) \mathcal{A}^{ij}\left( \frac{1}{4}h_{i}h_{j}+\frac{3}{8}\xi
\frac{\nu ^{2/3}}{\tilde{\nu}}h_{i}\widehat{d}_{j}+\left( \frac{3}{8}\right)
^{2}\xi ^{2}\frac{\nu ^{4/3}}{\tilde{\nu}^{2}}\widehat{d}_{i}\widehat{d}%
_{j}\right)
\end{array}
\right] \left( p^{0}\right) ^{2}+ \\
\\
+2\left[
\begin{array}{l}
-\frac{h}{12}h_{i}+\frac{1}{16}\xi \frac{\nu ^{2/3}}{\tilde{\nu}}h\widehat{d}%
_{i}-\left( \frac{3}{8}\right) ^{2}\xi ^{2}\frac{\nu ^{2}}{\tilde{\nu}^{2}}(%
\widehat{d}_{j}x^{j})\widehat{d}_{i}+ \\
\\
+\frac{3}{16}\xi \frac{\nu ^{2/3}}{\tilde{\nu}}(\widehat{d}_{j}x^{j})h_{i}+%
\tilde{\nu}^{2}\left( 1-\left( \frac{3}{4}\right) ^{2}\frac{\xi ^{2}}{\tilde{%
\nu}^{2}}\right) 12\mathcal{A}_{i}+ \\
\\
+\frac{1}{12}\left( 1-\left( \frac{3}{4}\right) ^{2}\frac{\xi ^{2}}{\tilde{%
\nu}^{2}}\right) \mathcal{A}^{kl}\left( \frac{1}{2}h_{ik}h_{l}+\frac{3}{8}%
\xi \frac{\nu ^{2/3}}{\tilde{\nu}}h_{ik}\widehat{d}_{l}\right)
\end{array}
\right] p^{0}p^{i}+ \\
\\
+\left[
\begin{array}{l}
\frac{1}{4}h_{i}h_{j}-\frac{3}{16}\xi \frac{\nu ^{2/3}}{\tilde{\nu}}(h_{i}%
\widehat{d}_{j}+h_{j}\widehat{d}_{i})+\left( \frac{3}{8}\right) ^{2}\xi ^{2}%
\frac{\nu ^{4/3}}{\tilde{\nu}^{2}}\widehat{d}_{i}\widehat{d}_{j}+ \\
\\
-\tilde{\nu}^{2}\left( 1-\left( \frac{3}{4}\right) ^{2}\frac{\xi ^{2}}{%
\tilde{\nu}^{2}}\right) 12\mathcal{A}_{ij}+ \\
\\
-\frac{1}{12}\left( 1-\left( \frac{3}{4}\right) ^{2}\frac{\xi ^{2}}{\tilde{%
\nu}^{2}}\right) \mathcal{A}^{kl}h_{ik}h_{jl}
\end{array}
\right] p^{i}p^{j}+ \\
\\
+\left( \frac{h}{3}-\frac{3}{4}\xi \frac{\nu ^{2/3}}{\tilde{\nu}}\widehat{d}%
_{i}x^{i}\right) q_{0}p^{0}+ \\
\\
+2\left[
\begin{array}{l}
\frac{h}{6}x^{i}-\frac{3}{8}\xi \frac{\nu ^{2/3}}{\tilde{\nu}}\widehat{d}%
_{j}x^{j}x^{i}-\frac{1}{24}\left( 1-\left( \frac{3}{4}\right) ^{2}\frac{\xi
^{2}}{\tilde{\nu}^{2}}\right) \mathcal{A}^{ij}h_{j}+ \\
\\
-\frac{3}{8}\cdot \frac{1}{12}\left( 1-\left( \frac{3}{4}\right) ^{2}\frac{%
\xi ^{2}}{\tilde{\nu}^{2}}\right) \xi \frac{\nu ^{2/3}}{\tilde{\nu}}\mathcal{%
A}^{ij}\widehat{d}_{j}
\end{array}
\right] q_{i}p^{0}+ \\
\\
+2\left( -\frac{h_{i}}{2}+\frac{3}{8}\xi \frac{\nu ^{2/3}}{\tilde{\nu}}%
\widehat{d}_{i}\right) q_{0}p^{i}+ \\
\\
+2\left[ -\frac{1}{2}h_{j}x^{i}+\frac{3}{8}\xi \frac{\nu ^{2/3}}{\tilde{\nu}}%
\widehat{d}_{j}x^{i}+\frac{1}{12}\left( 1-\left( \frac{3}{4}\right) ^{2}%
\frac{\xi ^{2}}{\tilde{\nu}^{2}}\right) \mathcal{A}^{ik}h_{jk}\right]
q_{i}p^{j}+ \\
\\
+q_{0}^{2}+ \\
\\
+2x^{i}q_{0}q_{i}+ \\
\\
+\left[ x^{i}x^{j}-\frac{1}{12}\left( 1-\left( \frac{3}{4}\right) ^{2}\frac{%
\xi ^{2}}{\tilde{\nu}^{2}}\right) \mathcal{A}^{ij}\right] q_{i}q_{j}
\end{array}
\right]
\end{array}
.  \notag \\
&&  \label{V_BH-d-SKG-lambda}
\end{eqnarray}

By using the results (\ref{1}) and (\ref{2}) of Appendix A, it is easy to
check that in the \textit{classical} limit $\xi \rightarrow 0$ Eq.(\ref
{V_BH-d-SKG-lambda}) yields the effective BH potential $\breve{V}_{BH}\left(
x^{j},\widehat{\lambda }^{j},\nu ;Q,0\right) $ for a generic $d$-SKG, given
by Eq. (2.13) of \cite{CFM1}, which we report here for ease of comparison:
\begin{eqnarray}
2\lim_{\xi \rightarrow 0}V_{BH} &=&2\breve{V}_{BH}=  \notag \\
&=&\left[ \nu \left( 1+4\breve{g}\right) +\frac{h^{2}}{36\nu }+\frac{3}{%
48\nu }\breve{g}^{ij}h_{i}h_{j}\right] \left( p^{0}\right) ^{2}+  \notag \\
&&+\left[ 4\nu \breve{g}_{ij}+\frac{1}{4\nu }\left( h_{i}h_{j}+\breve{g}%
^{mn}h_{im}h_{nj}\right) \right] p^{i}p^{j}+  \notag \\
&&+\frac{1}{\nu }\left[ q_{0}^{2}+2x^{i}q_{0}q_{i}+\left( x^{i}x^{j}+\frac{1%
}{4}\breve{g}^{ij}\right) q_{i}q_{j}\right] +  \notag \\
&&+2\left[ \nu \breve{g}_{i}-\frac{h}{12\nu }h_{i}-\frac{1}{8\nu }\breve{g}%
^{jm}h_{m}h_{ij}\right] p^{0}p^{i}+  \notag \\
&&-\frac{1}{3\nu }\left[ -hp^{0}q_{0}+3q_{0}p^{i}h_{i}-\left( hx^{i}+\frac{3%
}{4}\breve{g}^{ij}h_{j}\right) p^{0}q_{i}+3\left( h_{j}x^{i}+\frac{1}{2}%
\breve{g}^{im}h_{mj}\right) q_{i}p^{j}\right] .  \notag \\
&&  \label{V_BH-d-SKG}
\end{eqnarray}
$\breve{g}_{ij}$ and $\breve{g}^{ij}$ have been respectively defined in (\ref
{3}) and (\ref{4}) of Appendix A, with contractions consistently defined as
\begin{eqnarray}
\breve{g}_{i} &\equiv &\breve{g}_{ij}x^{j}=\lim_{\xi \rightarrow
0}g_{ij}x^{j}; \\
\breve{g} &\equiv &\breve{g}_{ij}x^{i}x^{j}=\lim_{\xi \rightarrow
0}g_{ij}x^{i}x^{j}.
\end{eqnarray}

\subsection{\label{Axion-Free-Supp-Configs}Axion-Free-Supporting
Configurations}

Let us now consider the terms of $V_{BH}$ given by Eq. (\ref
{V_BH-d-SKG-lambda}) which are linear in the \textit{axions} $x^{i}$'s; they
reads as follows:
\begin{eqnarray}
&&
\begin{array}{l}
\left. V_{BH}\right| _{linear~in~\left\{ x^{i}\right\} }=\frac{1}{2\tilde{\nu%
}}\left( 1-\left( \frac{3}{4}\right) ^{2}\frac{\xi ^{2}}{\tilde{\nu}^{2}}%
\right) ^{-1}\cdot \\
\\
\cdot \left[
\begin{array}{l}
2\left[
\begin{array}{l}
-\left( \frac{3}{8}\right) ^{2}\xi ^{2}\frac{\nu ^{4/3}}{\tilde{\nu}^{2}}(%
\widehat{d}_{j}x^{j})\widehat{d}_{i}+ \\
\\
+\tilde{\nu}^{2}\left( 1-\left( \frac{3}{4}\right) ^{2}\frac{\xi ^{2}}{%
\tilde{\nu}^{2}}\right) 12\mathcal{A}_{i}+ \\
\\
+\frac{3}{8}\cdot \frac{1}{12}\left( 1-\left( \frac{3}{4}\right) ^{2}\frac{%
\xi ^{2}}{\tilde{\nu}^{2}}\right) \xi \frac{\nu ^{2/3}}{\tilde{\nu}}\mathcal{%
A}^{kl}h_{ik}\widehat{d}_{l}
\end{array}
\right] p^{0}p^{i}+ \\
\\
-\frac{3}{4}\xi \frac{\nu ^{2/3}}{\tilde{\nu}}\widehat{d}_{i}x^{i}q_{0}p^{0}+
\\
\\
+2\left[ \frac{3}{8}\xi \frac{\nu ^{2/3}}{\tilde{\nu}}\widehat{d}_{j}x^{i}+%
\frac{1}{12}\left( 1-\left( \frac{3}{4}\right) ^{2}\frac{\xi ^{2}}{\tilde{\nu%
}^{2}}\right) \mathcal{A}^{ik}h_{jk}\right] q_{i}p^{j}+ \\
\\
\\
+2x^{i}q_{0}q_{i}
\end{array}
\right]
\end{array}
.  \notag \\
&&  \label{V_BH-d-SKG-lambda-linear-x}
\end{eqnarray}
As a consequence, for a $d$-SKG corrected by $\xi \neq 0$ (with prepotential
given by Eq. (\ref{prepot})), only two \textit{axion-free-supporting} BH
charge configuration exists, namely the \textit{electric} ($D2-D6$) and
\textit{magnetic} ($D0-D4$) ones:
\begin{eqnarray}
electric &:&Q_{el}\equiv \left(
\begin{array}{c}
p^{0} \\
0 \\
0 \\
q_{i}
\end{array}
\right) ;  \label{Q-el} \\
magnetic &:&Q_{magn}\equiv \left(
\begin{array}{c}
0 \\
p^{i} \\
q_{0} \\
0
\end{array}
\right) .  \label{Q-magn}
\end{eqnarray}
For such BH charge configurations $x^{i}=0$ $\forall i$ is a(n \textit{at
least)} particular solution of the axionic Attractor Eqs.:
\begin{eqnarray}
\left. \frac{\partial V_{BH}}{\partial x^{i}}\right| _{Q=Q_{el}}
&=&0\Leftarrow x^{i}=0~\forall i;  \label{a-1} \\
\left. \frac{\partial V_{BH}}{\partial x^{i}}\right| _{Q=Q_{magn}}
&=&0\Leftarrow x^{i}=0~\forall i.  \label{a-2}
\end{eqnarray}
This fact is a major difference with respect to the classical limit $\xi
\rightarrow 0$, in which the linear term in $x^{i}$'s proportional to $%
q_{0}p^{0}$ (see Eq. (\ref{V_BH-d-SKG-lambda-linear-x})) vanishes. Indeed,
it consistently holds that
\begin{equation}
2\lim_{\xi \rightarrow 0}\left. V_{BH}\right| _{linear~in~\left\{
x^{i}\right\} }=2\left. \breve{V}_{BH}\right| _{linear~in~\left\{
x^{i}\right\} }=\frac{1}{\nu }2x^{i}q_{0}q_{i}+2\nu \breve{g}_{i}p^{0}p^{i}-%
\frac{1}{2\nu }\breve{g}^{ik}h_{kj}q_{i}p^{j}.
\end{equation}
This implies that also the Kaluza-Klein ($D0-D6$) BH charge configuration
\begin{equation}
KK:Q_{KK}\equiv \left(
\begin{array}{c}
p^{0} \\
0 \\
q_{0} \\
0
\end{array}
\right)
\end{equation}
supports \textit{axion-free} (\textit{at least} particular) attractor
solutions \cite{CFM1}. Thus, besides the classical limits of Eqs. (\ref{a-1}%
) and (\ref{a-2}), namely:
\begin{eqnarray}
\left. \frac{\partial \breve{V}_{BH}}{\partial x^{i}}\right| _{Q=Q_{el}}
&=&0\Leftarrow x^{i}=0~\forall i; \\
\left. \frac{\partial \breve{V}_{BH}}{\partial x^{i}}\right| _{Q=Q_{magn}}
&=&0\Leftarrow x^{i}=0~\forall i,
\end{eqnarray}
for $\xi =0$ it also holds that
\begin{equation}
\left. \frac{\partial \breve{V}_{BH}}{\partial x^{i}}\right|
_{Q=Q_{KK}}=0\Leftarrow x^{i}=0~\forall i.
\end{equation}

The \textit{non-axion-free-supporting} nature of the $D0-D6$ BH charge
configuration in perturbatively quantum corrected $d$-SKG (determined by the
holomorphic prepotential (\ref{prepot})) is consistent with, and sheds new
light on, the results of \cite{BFMS2}.

Such a paper (developing the analysis of \cite{BFMS1}) addressed the issue
of the fate of the unique non-BPS $Z\neq 0$ \textit{flat} direction in the $%
\mathcal{N}=2$, $d=4$ \textit{ungauged} Maxwell-Einstein supergravity
described by Eq. (\ref{prepot}) with $n_{V}=2$ (\textit{i.e.} the so-called
``$st^{2}+i\xi $ model''). By analyzing the (supergravity analogues of the) $%
D0-D4$, $D2-D6$ and $D0-D6$ charge configurations, the following results
were obtained:

\begin{itemize}
\item  In $D0-D4$ and $D2-D6$ charge configurations the classical solutions (%
$\xi =0$) were found to lift at the quantum level ($\xi \neq 0$).
Remarkably, it was found that the \textit{quantum} \textit{lift} occurs more
often towards \textit{repeller} directions (thus destabilizing the whole
critical solution, and \textit{destroying the attractor in strict sense}),
rather than towards \textit{attractor} directions.

\item  The $D0-D6$ charge configuration yielded a somewhat surprising
result: the classical solution gets modified at the \textit{quantum} level,
acquiring a \textit{non-vanishing} axionic part. However, despite being no
more purely imaginary, \textit{such a quantum non-BPS }$Z\neq 0$\textit{\
solution still exhibits a flat direction}. The origin of such a deep
difference among \textit{electric/magnetic} and $D0-D6$ configurations was
unclear in \cite{BFMS2}, but it is clarified (and further generalized to an
arbitrary number $n_{V}$ of Abelian vector multiplets) from the results of
the analysis performed above: due to the very structure of $V_{BH}$ (see
Eqs. (\ref{V_BH-d-SKG-lambda} and (\ref{V_BH-d-SKG-lambda-linear-x})) for $%
\xi \neq 0$, the \textit{electric/magnetic }still support axion-free
solutions, whereas the $D0-D6$ configuration do \textit{not}.
\end{itemize}

On the other hand, the persistence of the flat direction also in presence of
\textit{quantum generated axions} is still not completely understood, and we
left the study of such issues for future work.

\section{\label{Z-and-DZ}Central Charge and \textit{Matter Charges}}

As given by Eq. (\ref{VBH1}), the effective BH potential $V_{BH}$ enjoys a
rewriting in terms of the $\mathcal{N}=2$, $d=4$ central charge $Z$ and of
its covariant derivatives $D_{i}Z$ (usually named \textit{matter charges}),
which is therefore worth computing.

In order to do this, let us recall that under the assumptions (\ref
{special-coords})-(\ref{Kgf}) the holomorphic prepotential (\ref{prepot})
reduces to
\begin{equation}
\mathcal{F}(z;\xi )\equiv \frac{1}{3!}d_{ijk}z^{i}z^{j}z^{k}+i\xi .
\label{effe}
\end{equation}
Furthermore, the K\"{a}hler potential reads ($\mathcal{F}_{i}\equiv \partial
\mathcal{F}/\partial z^{i}$; see \textit{e.g.} \cite{CDF-review,FerBig})
\begin{eqnarray}
K &=&-\log \left\{ i\left[ 2(\mathcal{F}-\overline{\mathcal{F}})+(\overline{z%
}^{\overline{\imath }}-z^{i})(\mathcal{F}_{i}+\overline{\mathcal{F}}_{%
\overline{\imath }})\right] \right\} =  \notag \\
&=&-\log \left[ -\frac{i}{3!}d_{ijk}(z^{i}-\overline{z}^{\overline{\imath }%
})(z^{j}-\overline{z}^{\overline{\jmath }})(z^{k}-\overline{z}^{\overline{k}%
})-4\xi \right] =  \notag \\
&=&-\log \left( 8\left( \nu -\frac{\xi }{2}\right) \right) ,  \label{K}
\end{eqnarray}
where definitions (\ref{z-split-def}) and (\ref{nu}) were used. Eq. (\ref{K}%
) thus implies
\begin{equation}
\exp \left( -K\right) =8\left( \nu -\frac{\xi }{2}\right) \Leftrightarrow
\exp \left( K/2\right) =\frac{1}{2\sqrt{2\nu -\xi }},
\end{equation}
with the global condition of consistency (relevant also for previous
treatment, see for instance Eq. (\ref{covmetric}))
\begin{equation}
2\nu -\xi >0.
\end{equation}

Therefore, by recalling its very definition (see \textit{e.g.} \cite{FGK}
and Refs. therein)
\begin{equation}
Z\equiv e^{K/2}(X^{\Lambda }q_{\Lambda }-F_{\Lambda }p^{\Lambda })\equiv
e^{K/2}W,
\end{equation}
where $W$ is the holomorphic superpotential, and under the assumptions (\ref
{special-coords})-(\ref{Kgf}), the $\mathcal{N}=2$, $d=4$ central charge
function for the holomorphic prepotential (\ref{prepot}) can be computed to
be:
\begin{eqnarray}
Z\left( x^{j},\widehat{\lambda }^{j},\nu ;Q,\xi \right) &=&\frac{1}{2\sqrt{%
2\nu -\xi }}W\left( x^{j},\widehat{\lambda }^{j},\nu ;Q,\xi \right) =  \notag
\\
&&  \notag \\
&=&\frac{1}{2\sqrt{2\nu -\xi }}\left[
\begin{array}{l}
q_{0}+q_{i}x^{i}-\frac{p^{0}}{2}\nu ^{2/3}\widehat{d}_{i}x^{i}+\frac{p^{0}}{6%
}h-\frac{p^{i}}{2}h_{i}+\nu ^{2/3}\frac{p^{i}}{2}\widehat{d}_{i}+ \\
\\
+i\nu ^{1/3}\left( -q_{i}\widehat{\lambda }^{i}-\frac{p^{0}}{2}\widehat{d}%
_{ij}x^{i}x^{j}+p^{0}\nu ^{2/3}-2\frac{\xi }{\nu ^{1/3}}p^{0}+p^{i}\widehat{d%
}_{ij}x^{j}\right)
\end{array}
\right] ;  \notag \\
&&  \label{Z} \\
D_{i}Z\left( x^{j},\widehat{\lambda }^{j},\nu ;Q,\xi \right) &=&\frac{1}{2%
\sqrt{2\nu -\xi }}D_{i}W\left( x^{j},\widehat{\lambda }^{j},\nu ;Q,\xi
\right) =  \notag \\
&&  \notag \\
&=&\frac{1}{2\sqrt{2\nu -\xi }}\left\{
\begin{array}{l}
q_{i}+\frac{p^{0}}{2}h_{i}-\frac{p^{0}}{4}\nu ^{2/3}\widehat{d}_{i}+ \\
-p^{j}h_{ij}+i\nu ^{1/3}\left( -p^{0}x^{j}+p^{j}\right) \widehat{d}_{ij}+ \\
\\
-\frac{i}{2}\frac{\nu ^{2/3}}{(2\nu -\xi )}\widehat{d}_{i}\left[
\begin{array}{l}
q_{0}+q_{j}x^{j}-\frac{p^{0}}{2}\nu ^{2/3}\widehat{d}_{j}x^{j}+\frac{p^{0}}{6%
}h+ \\
-\frac{p^{j}}{2}h_{j}+\frac{p^{j}}{2}\nu ^{2/3}\widehat{d}_{j}+ \\
+i\nu ^{1/3}\left(
\begin{array}{l}
-q_{j}\widehat{\lambda }^{j}-\frac{p^{0}}{2}\widehat{d}_{jk}x^{j}x^{k}+ \\
+p^{0}\nu ^{2/3}-2\frac{\xi }{\nu ^{1/3}}p^{0}+p^{j}\widehat{d}_{jk}x^{k}
\end{array}
\right)
\end{array}
\right]
\end{array}
\right\} .  \notag \\
&&  \label{DZ}
\end{eqnarray}
Clearly, due to the different K\"{a}hler weights of $Z$ and $W$
(respectively $\left( 1,-1\right) $ and $\left( 2,0\right) $), the covariant
differential operator acting on them has different definitions, namely:
\begin{eqnarray}
D_{i}Z &\equiv &\partial _{i}Z+\frac{1}{2}\left( \partial _{i}K\right) Z; \\
D_{i}W &\equiv &\partial _{i}W+\left( \partial _{i}K\right) W.
\end{eqnarray}

Notice that in the limit $\xi \rightarrow 0$ Eqs. (\ref{Z}) and (\ref{DZ})
exactly matches with known results for $d$-SK geometries, given by Eq.(4.9)
and (4.10) of \cite{CFM1}. It is worth remarking that Eq. (\ref{Z}) yields
that the holomorphic superpotential $W$ gets modified, with respect to its
classical ($\xi \rightarrow 0$) counterpart, only by a \textit{global} shift
of its imaginary part:
\begin{equation}
W\left( x^{j},\widehat{\lambda }^{j},\nu ;Q,\xi \right) =W\left( x^{j},%
\widehat{\lambda }^{j},\nu ;Q,0\right) -2\xi ip^{0}.
\end{equation}
In particular, for \textit{axion-free} critical solutions (supported for $%
\xi \neq 0$ only by the charge configurations (\ref{Q-el}) and (\ref{Q-magn}%
)) it holds that the superpotential $W$ (\textit{on-shell} for axions $x^{i}$%
's) is purely imaginary and real, respectively:
\begin{eqnarray}
W\left( x^{j}=0,\widehat{\lambda }^{j},\nu ;Q_{el},\xi \right) &=&i\nu
^{1/3}\left( -q_{i}\widehat{\lambda }^{i}+p^{0}\nu ^{2/3}-2\frac{\xi }{\nu
^{1/3}}p^{0}\right) ; \\
W\left( x^{j}=0,\widehat{\lambda }^{j},\nu ;Q_{magn},\xi \right)
&=&q_{0}+\nu ^{2/3}\frac{p^{i}}{2}\widehat{d}_{i}.
\end{eqnarray}
\medskip

Concerning supersymmetric critical points of $V_{BH}$, the ($\frac{1}{2}$%
-)BPS conditions
\begin{equation}
D_{i}W=0~\forall i=1,...,n_{V}  \label{1/2-BPS-conds}
\end{equation}
for \textit{axion-free} critical solutions within the charge configurations (%
\ref{Q-el}) and (\ref{Q-magn}) respectively read ($\forall i=1,...,n_{V}$):
\begin{eqnarray}
D_{i}W\left( x^{j}=0,\widehat{\lambda }^{j},\nu ;Q_{el},\xi \right)
&=&0\Leftrightarrow q_{i}-\frac{p^{0}}{4}\nu ^{2/3}\widehat{d}_{i}+\frac{1}{2%
}\frac{\nu }{(2\nu -\xi )}\widehat{d}_{i}\left( -q_{j}\widehat{\lambda }%
^{j}+p^{0}\nu ^{2/3}-2\frac{\xi }{\nu ^{1/3}}p^{0}\right) =0;  \notag \\
&& \\
D_{i}W\left( x^{j}=0,\widehat{\lambda }^{j},\nu ;Q_{magn},\xi \right)
&=&0\Leftrightarrow p^{j}\widehat{d}_{ij}-\frac{1}{2}\frac{\nu ^{1/3}}{(2\nu
-\xi )}\widehat{d}_{i}\left( q_{0}+\frac{p^{j}}{2}\nu ^{2/3}\widehat{d}%
_{j}\right) =0,
\end{eqnarray}
reducing to $n_{V}$ ($\xi $-parametrized) real algebraic Eqs. in $n_{V}$
real unknowns $\left\{ \widehat{\lambda }^{i},\nu \right\} $.

In \cite{CQ-N=2-BHs}\ the \textit{axion-free} $\frac{1}{2}$-BPS critical
points of $V_{BH}$ determined by the holomorphic prepotential (\ref{effe})
were determined by introducing the K\"{a}hler gauge-invariant sections
\begin{equation}
Y\equiv \overline{Z}\left(
\begin{array}{c}
L^{0} \\
L^{i} \\
M_{0} \\
M_{i}
\end{array}
\right) =\exp \left( K\right) \overline{W}\left(
\begin{array}{c}
X^{0} \\
X^{i} \\
F_{0} \\
F_{i}
\end{array}
\right)
\end{equation}
and evaluating the identities of the SK geometries (see \textit{e.g. }\cite
{CDF-review,FerBig} and Refs. therein\textit{)} along the BPS conditions (%
\ref{1/2-BPS-conds}), thus obtaining ($\Xi \in \mathbb{R}$)
\begin{equation}
\left\{
\begin{array}{l}
Y^{0}=\frac{1}{2}\left( \Xi +ip^{0}\right) ; \\
~ \\
Y^{i}=ip^{i}\frac{\left( \Xi +ip^{i}\right) }{\Xi }.
\end{array}
\right.
\end{equation}
In the case of $\Xi \neq 0$, the $\xi $-dependent value of $V_{BH}$ at its $%
\frac{1}{2}$-BPS \textit{axion-free} critical points can be computed to be%
\footnote{%
Eq. (\ref{V_BPS-axion-free-CSI<>0}) fixes a typo in Eq. (3.35) of \cite
{CQ-N=2-BHs}. For the configuration $D0-D6$ (which however, as explicitly
shown above, is \textit{not axion-free-supporting} for $\xi \neq 0$) this
was noticed in \cite{BFMS1}.}
\begin{equation}
V_{BH,BPS,axion-free}=-2\left[ \Xi +\frac{\left( p^{0}\right) ^{2}}{\Xi }%
\right] \left( q_{0}-2\xi \Xi +\frac{\xi }{2}\Xi \right) ,
\label{V_BPS-axion-free-CSI<>0}
\end{equation}
where $\Xi $ satisfies the $\xi $-parametrized Eq. (see Eq. (3.34) of \cite
{CQ-N=2-BHs}):
\begin{equation}
3p^{0}q_{0}+p^{i}q_{i}=6\xi \Xi p^{0},
\end{equation}
along with the condition \cite{CQ-N=2-BHs}
\begin{equation}
\left( q_{0}-2\Xi \right) d_{ijk}p^{i}p^{j}p^{k}>0.
\end{equation}
On the other hand, the $\frac{1}{2}$-BPS \textit{axion-free} solutions with $%
\Xi =0$ are necessarily supported only by the \textit{electric}
configuration (\ref{Q-el}), and the dependence on $\xi $ drops out: the
resolution of the Attractor Eqs. in terms of the sections $Y^{\Lambda }$'s
and the determination of the critical value of $V_{BH}$ go as for a generic $%
d$-SK geometry \cite{Shmakova}.\medskip

Concerning \textit{non-axion-free} supersymmetric (\textit{if any}) and
non-supersymmetric (either \textit{axion-free} or \textit{non-axion-free})
critical points of $V_{BH}$, the case study becomes much more complicated.

As yielded by the analysis of $t^{3}+i\xi $ model ($n_{V}=1$) \cite{BFMS1}
and of $st^{2}+i\xi $ model ($n_{V}=2$ particular case) \cite{BFMS2}, in
general the corresponding Attractor Eqs. are higher-order algebraic Eqs.
which cannot be solved analytically, but only investigated numerically.
Furthermore, interesting phenomena occur, such as: the \textit{``separation''%
} of attractors (related to presence of \textit{basins of attraction }/
\textit{area codes} in the dynamical system describing the radial evolution
of the scalar fields in the BH space-time background) \cite{BFMS1}; the
\textit{``transmutation''} of the supersymmetry-preserving properties of the
attractors \cite{BFMS1}; and the \textit{``lifting''} (with or without
removal) of the \textit{``flat''} directions of the critical potential,
which exist in the classical ($\xi =0$) regime, at least for \textit{%
symmetric} $d$-SK geometries \cite{BFMS2}.

Despite the lack of analytical expressions of non-supersymmetric (non-BPS $%
Z\neq 0$ \textit{and/or} non-BPS $Z=0$) critical points of $V_{BH}$ for $\xi
\neq 0$, many issues are still to be carefully investigated (we list some of
them in the concluding Sect. \ref{Conclusion}).

The intricacy of the SK geometry described by the holomorphic prepotential (%
\ref{prepot}) (or, equivalently by Eq. (\ref{effe})) calls for a deeper
analysis of the fundamental quantities characterizing such a geometry, and
also for a deeper understanding of the conditions determining the (various
classes of) critical points of $V_{BH}$ itself. The study of these issues,
needed for a deeper investigation of the dynamics of the \textit{Attractor
Mechanism} in the generally non-homogeneous geometries under consideration,
will be the object of Sects. \ref{E-Tensor} and \ref{Riemann-Tensor}.

\section{\label{E-Tensor}$E$-tensor}

The first quantity we want to determine is the so-called $E$-tensor. This
rank-$5$ tensor was firstly introduced in \cite{dWVVP} (see also the
treatment of \cite{CVP}), and it expresses the deviation of the considered
geometry from being symmetric. Its definition reads (see \textit{e.g.} \cite
{Kallosh-review} for a recent treatment, and Refs. therein):
\begin{equation}
\overline{E}_{\overline{m}ijkl}\equiv \frac{1}{3}\overline{D}_{\overline{m}%
}D_{i}C_{jkl}.  \label{E}
\end{equation}
This definition can be elaborated further, by recalling the properties of
the so-called $C$-tensor $C_{ijk}$. This is a rank-$3$ tensor with
K\"{a}hler weights $\left( 2,-2\right) $, defined as (see \textit{e.g.} \cite
{CDF-review,Castellani1,DFF}):
\begin{eqnarray}
C_{ijk} &\equiv &\left\langle D_{i}D_{j}V,D_{k}V\right\rangle =e^{K}\left(
\partial _{i}\mathcal{N}_{\Lambda \Sigma }\right) D_{j}X^{\Lambda
}D_{k}X^{\Sigma }=  \notag \\
&=&e^{K}\left( \partial _{i}X^{\Lambda }\right) \left( \partial
_{j}X^{\Sigma }\right) \left( \partial _{k}X^{\Xi }\right) \partial _{\Xi
}\partial _{\Sigma }F_{\Lambda }\left( X\right) \equiv  \notag \\
&\equiv &e^{K}W_{ijk},\text{~~}\overline{\partial }_{\overline{l}}W_{ijk}=0,
\label{C}
\end{eqnarray}
where the second line holds only in \textit{special coordinates}. $C_{ijk}$
is completely symmetric and covariantly holomorphic:
\begin{eqnarray}
C_{ijk} &=&C_{\left( ijk\right) }; \\
\overline{D}_{\overline{i}}C_{jkl} &=&0.
\end{eqnarray}
Furthermore, it enters the fundamental constraints on the Riemann tensor $%
R_{i\overline{j}k\overline{l}}$ of SK geometry \footnote{%
Notice that the third of Eqs. (\ref{C}) correctly defines the Riemann tensor
$R_{i\overline{j}k\overline{l}}$, and it is actual the opposite of the one
which may be found in a large part of existing literature. Indeed, such a
formulation yields negative values of the constant scalar curvature
homogeneous symmetric non-compact SK manifolds, as given by the treatment of
\cite{CVP}.} (see \textit{e.g.} \cite{CDF-review,Castellani1,DFF}, \cite
{CDF-review} and Refs. therein; see also \textit{e.g.} \cite{Valencia} and
\cite{Kallosh-review} for more recent reviews):
\begin{equation}
R_{i\overline{j}k\overline{l}}=-g_{i\overline{j}}g_{k\overline{l}}-g_{i%
\overline{l}}g_{k\overline{j}}+C_{ikm}\overline{C}_{\overline{j}\overline{l}%
\overline{n}}g^{m\overline{n}}.  \label{SKG-constraints}
\end{equation}
The Bianchi identities for $R_{i\overline{j}k\overline{l}}$ (see \textit{e.g.%
} \cite{Castellani1}) and constraints (\ref{SKG-constraints}) yield the
following result
\begin{equation}
D_{[i}C_{j]kl}=0,
\end{equation}
where (round) square brackets denote (symmetrization) anti-symmetrization
with respect to enclosed indices throughout. Due to its holomorphic
K\"{a}hler weight, the covariant derivative of $C_{ijk}$ reads:
\begin{equation}
D_{i}C_{jkl}=D_{(i}C_{j)kl}=\partial _{i}C_{jkl}+\left( \partial
_{i}K\right) C_{jkl}+\Gamma _{ij}^{~~m}C_{mkl}+\Gamma
_{ik}^{~~m}C_{mjl}+\Gamma _{il}^{~~m}C_{mjk},  \label{DC}
\end{equation}
where the Christoffel connection $\Gamma $ is defined as
\begin{equation}
\Gamma _{ij}^{~~m}\equiv -g^{m\overline{l}}\partial _{i}g_{j\overline{l}}.
\label{Gamma}
\end{equation}
By using Eqs. (\ref{C})-(\ref{Gamma}), $\overline{E}_{\overline{m}ijkl}$
defined by (\ref{E}) can thus be further elaborated as follows:
\begin{eqnarray}
\overline{E}_{\overline{m}ijkl} &=&\frac{1}{3}\overline{D}_{\overline{m}%
}D_{(i}C_{jkl)}=C_{p(kl}C_{ij)n}g^{n\overline{n}}g^{p\overline{p}}\overline{C%
}_{\overline{n}\overline{p}\overline{m}}-\frac{4}{3}g_{\left( l\right|
\overline{m}}C_{\left| ijk\right) }=  \notag \\
&=&g^{n\overline{n}}R_{\left( i\right| \overline{m}\left| j\right| \overline{%
n}}C_{n\left| kl\right) }+\frac{2}{3}g_{\left( i\right| \overline{m}%
}C_{\left| jkl\right) }=\overline{E}_{\overline{m}\left( ijkl\right) }.
\label{E-elab}
\end{eqnarray}
It thus holds that $\overline{E}_{\overline{m}ijkl}=0$ \textit{globally} in
(homogeneous) symmetric SK manifolds, defined by the covariant constancy of $%
R_{i\overline{j}k\overline{l}}$ itself:
\begin{equation}
D_{m}R_{i\overline{j}k\overline{l}}=0.  \label{DR=0}
\end{equation}
Eq. (\ref{DR=0}), through the covariant holomorphicity of $C_{ijk}$ and the
constraints (\ref{SKG-constraints}), yields the \textit{global} covariant
constancy of $C_{ijk}$ itself, and thus the global vanishing of $\overline{E}%
_{\overline{m}ijkl}$: \textbf{\ }
\begin{equation}
D_{i}C_{jkl}=D_{(i}C_{j)kl}=0\Rightarrow \overline{E}_{\overline{m}ijkl}=0,
\label{E=0}
\end{equation}
which in turn, through Eq. (\ref{E-elab}), implies
\begin{equation}
C_{p(kl}C_{ij)n}g^{n\overline{n}}g^{p\overline{p}}\overline{C}_{\overline{n}%
\overline{p}\overline{m}}=\frac{4}{3}g_{\left( l\right| \overline{m}%
}C_{\left| ijk\right) }\Leftrightarrow g^{n\overline{n}}R_{\left( i\right|
\overline{m}\left| j\right| \overline{n}}C_{n\left| kl\right) }=-\frac{2}{3}%
g_{\left( i\right| \overline{m}}C_{\left| jkl\right) }.  \label{symm}
\end{equation}
It is worth noticing that, while (\ref{DR=0}) defines the symmetricity of a
K\"{a}hler manifold, Eq. (\ref{E=0}) (or equivalently Eq. (\ref{symm})) is a
necessary (but not necessarily sufficient) condition of symmetricity.

Recently, in \cite{ADFT-review} the $E$-tensor was used in the expression of
the value of $V_{BH}$ at its non-BPS $Z\neq 0$ critical points (see also the
treatment in \cite{CFMZ1}, and Refs. therein):
\begin{equation}
V_{BH,nBPS,Z\neq 0}=\left[ 4\left| Z\right| ^{2}+\Delta \right] _{nBPS,Z\neq
0},  \label{V-1}
\end{equation}
where ($Z^{\overline{i}}\equiv g^{j\overline{i}}D_{j}Z$)
\begin{equation}
\Delta \equiv -\frac{3}{4}\frac{E_{m\overline{i}\overline{j}\overline{k}%
\overline{l}}Z^{\overline{i}}Z^{\overline{j}}Z^{\overline{k}}Z^{\overline{l}}%
\overline{Z}^{m}}{\left( \overline{C}_{\overline{n}\overline{p}\overline{q}%
}Z^{\overline{n}}Z^{\overline{p}}Z^{\overline{q}}\right) },  \label{Delta}
\end{equation}
such that (see \textit{e.g. }\cite{ADFT-review,Kallosh-review}).
\begin{gather}
\left[ \frac{g^{i\overline{j}}\left( D_{i}Z\right) \overline{D}_{\overline{j}%
}\overline{Z}}{\left| Z\right| ^{2}}\right] _{nBPS,Z\neq 0}=3;
\label{rule-of-3} \\
\Updownarrow  \notag \\
\Delta _{nBPS,Z\neq 0}=0\Leftrightarrow \left( E_{m\overline{i}\overline{j}%
\overline{k}\overline{l}}Z^{\overline{i}}Z^{\overline{j}}Z^{\overline{k}}Z^{%
\overline{l}}\overline{Z}^{m}\right) _{nBPS,Z\neq 0}=0,  \label{rule-of-3-1}
\end{gather}
where in the last step the non-degeneracy of the cubic norm $C_{ijk}%
\overline{Z}^{i}\overline{Z}^{j}\overline{Z}^{k}$ (\textit{at least} at
non-BPS $Z\neq 0$ critical points of $V_{BH}$) was used. Therefore, Eq. (\ref
{E=0}) (or equivalently Eq. (\ref{symm})) is a sufficient (but not
necessary) condition for the so-called \textit{``rule of three''} (\ref
{rule-of-3}) to hold at non-BPS $Z\neq 0$ critical points of $V_{BH}$.

These results (and further relations with the sectional curvature treated
further below; see Eqs. (\ref{Pp-1})-(\ref{UCLA11}) as well as the treatment
given in \cite{Kallosh-review}) call for an explicit determination of the $E$%
-tensor in the SK geometries described by the holomorphic prepotential (\ref
{effe}), and through the limit $\xi \rightarrow 0$, in a generic $d$-SK
geometry.

Thus, by a long but straightforward algebra (detailed in Appendix B), the
covariant derivative of the $C$-tensor can be written as follows:
\begin{equation}
D_{i}C_{jkl}=\frac{i}{2^{5}}\frac{1}{\left( \nu -\frac{\xi }{2}\right) ^{2}}%
\left[
\begin{array}{l}
-\frac{\left( \nu -\frac{\xi }{2}\right) }{\nu +\xi }\nu ^{2/3}\left(
d_{ij}d_{kl}+d_{ik}d_{jl}+d_{il}d_{jk}\right) + \\
-2\left( \nu -\frac{\xi }{2}\right) \nu ^{-1/3}\left(
d_{ijn}d_{mkl}+d_{ikn}d_{mjl}+d_{iln}d_{mjk}\right) \widehat{d}^{mn}+ \\
+\nu ^{2/3}\left( d_{i}d_{jkl}+d_{j}d_{ikl}+d_{k}d_{ijl}+d_{l}d_{ijk}\right)
\end{array}
\right] .  \label{covderC}
\end{equation}
Notice that for $\xi \neq 0$ there is no way to make $D_{i}C_{jkl}=0$
\textit{globally}. This confirms the result of \cite{CVP} that, with the
exception of the sequence of the \textit{minimal coupling} sequence $\mathbb{%
C}\mathbb{P}^{n}$, all homogeneous symmetric (non-compact) SK are given by $%
d $-geometries (namely, by the $\xi \rightarrow 0$ limit of prepotential (%
\ref{effe})). Thus, the SK geometries described by the holomorphic
prepotential (\ref{effe}) are not symmetric, nor homogeneous (\textit{at
least} of the $d$-type studied and classified in \cite{CVP,dWVP,dWVVP}, and
Refs. therein).

Through definition (\ref{E}) and Eq. (\ref{covderC}), the $E$-tensor can
then be explicitly computed:
\begin{eqnarray}
\overline{E}_{\overline{m}ijkl} &\equiv &\frac{1}{3}\overline{D}_{\overline{m%
}}D_{i}C_{jkl}=\frac{1}{3}\left[ \overline{\partial }_{\overline{m}%
}D_{i}C_{jkl}-(\overline{\partial }_{\overline{m}}K)D_{i}C_{jkl}\right] =
\notag \\
&&  \notag \\
&=&\frac{1}{12\cdot 2^{4}}\frac{1}{\left( \nu -\frac{\xi }{2}\right) ^{2}}%
\left[
\begin{array}{l}
\left( 2\nu -7\xi \right) \frac{\nu ^{4/3}}{4\left( \nu +\xi \right) ^{2}}%
\left( \widehat{d}_{ij}\widehat{d}_{kl}+\widehat{d}_{ik}\widehat{d}_{jl}+%
\widehat{d}_{il}\widehat{d}_{jk}\right) \widehat{d}_{m}+ \\
\\
-\frac{\nu ^{4/3}}{2\left( \nu -\frac{\xi }{2}\right) }\left( \widehat{d}%
_{i}d_{jkl}+\widehat{d}_{j}d_{ikl}+\widehat{d}_{k}d_{ijl}+\widehat{d}%
_{l}d_{ijk}\right) \widehat{d}_{m}+ \\
\\
-\frac{\left( \nu -\frac{\xi }{2}\right) }{\nu +\xi }\nu ^{1/3}\left(
\begin{array}{l}
d_{ijm}\widehat{d}_{kl}+d_{klm}\widehat{d}_{ij}+d_{ikm}\widehat{d}_{jl}+ \\
+d_{jlm}\widehat{d}_{ik}+d_{ilm}\widehat{d}_{jk}+d_{jkm}\widehat{d}_{il}
\end{array}
\right) + \\
\\
+2\nu ^{1/3}\left( \widehat{d}_{im}d_{jkl}+\widehat{d}_{jm}d_{ikl}+\widehat{d%
}_{km}d_{ijl}+\widehat{d}_{lm}d_{ijk}\right) + \\
\\
-2\left( \nu -\frac{\xi }{2}\right) \left(
d_{ijn}d_{pkl}+d_{ikn}d_{pjl}+d_{iln}d_{pjk}\right) \frac{\partial (\tilde{d}%
^{-1})^{pn}}{\partial \lambda ^{m}}
\end{array}
\right] ,  \notag \\
&&  \label{E-csi-1}
\end{eqnarray}

where it is easy to show that
\begin{equation}
\frac{\partial d^{pn}}{\partial \lambda ^{m}}=-d_{ijm}d^{ip}d^{jn}=-\nu
^{-2/3}d_{ijm}\widehat{d}^{ip}\widehat{d}^{jn}.  \label{ex1}
\end{equation}

By standard symmetrization procedures and using Eq. (\ref{ex1}), Eq. (\ref
{E-csi-1}) can be further elaborated as follows:
\begin{equation}
\overline{E}_{\overline{m}ijkl}=-\frac{1}{3\cdot 2^{7}}\frac{1}{\left( \nu -%
\frac{\xi }{2}\right) ^{3}}\left[
\begin{array}{l}
\left[ 4\;\widehat{d}_{(i}d_{jkl)}-3\;\frac{\left( \nu -\frac{\xi }{2}%
\right) }{\nu +\xi }\;\widehat{d}_{(ij}\widehat{d}_{kl)}\right] \nu ^{4/3}%
\widehat{d}_{m}+ \\
\\
+12\;\frac{\left( \nu -\frac{\xi }{2}\right) ^{2}}{\nu +\xi }\;\nu
^{1/3}d_{m(ij}\widehat{d}_{kl)}-2^{4}\left( \nu -\frac{\xi }{2}\right) \nu
^{1/3}\;\widehat{d}_{m(i}d_{jkl)}+ \\
\\
-12\left( \nu -\frac{\xi }{2}\right) ^{2}\;\nu
^{-2/3}d_{p(ij}d_{kl)n}\;d_{mrs}\widehat{d}^{rp}\widehat{d}^{sn}+ \\
\\
+\frac{3}{2}\xi \;\frac{\left( \nu -\frac{\xi }{2}\right) }{\left( \nu +\xi
\right) ^{2}}\nu ^{4/3}\;\widehat{d}_{m}\;\widehat{d}_{(ij}\widehat{d}_{kl)}
\end{array}
\right] .  \label{ETfinal}
\end{equation}

It is here worth remarking that the observation made above that for $\xi
\neq 0$ it is not possible to make $D_{i}C_{jkl}=0$ \textit{globally} does
not imply that $\overline{E}_{\overline{m}ijkl}=0$, \textit{and/or} $E_{m%
\overline{i}\overline{j}\overline{k}\overline{l}}Z^{\overline{i}}Z^{%
\overline{j}}Z^{\overline{k}}Z^{\overline{l}}\overline{Z}^{m}=0$, \textit{%
locally}, namely on a (set of) point(s), eventually at non-BPS $Z\neq 0$
critical points of $V_{BH}$. Thus, the interesting question arises (which we
leave for future investigation) whether for some charge configurations (and
eventually for some value(s) of $\xi $ itself) the \textit{``rule of three''}
(\ref{rule-of-3}) still holds at non-BPS $Z\neq 0$ critical points of $%
V_{BH} $ in SK geometries determined by the prepotential (\ref{effe}). Let
us here recall that, as explicitly found in \cite{DFT-hom-non-symm}, \textit{%
at least} in some homogeneous \textit{non-symmetric} $d$-SK geometries, the
\textit{``rule of three''} (\ref{rule-of-3}) still holds, despite the fact
that $\overline{E}_{\overline{m}ijkl}$ does \textit{not} vanish globally.

Before concluding this Section, let us notice that in the limit $\xi
\rightarrow 0$ the result (\ref{ETfinal}) yields the expression of the $E$%
-tensor for a generic $d$-SK geometry, namely:
\begin{equation}
\overline{E}_{\overline{m}ijkl,\xi =0}=-\frac{1}{3\cdot 2^{7}}\nu ^{-5/3}%
\left[
\begin{array}{l}
\left( 4\;\widehat{d}_{(i}d_{jkl)}-3\;\widehat{d}_{(ij}\widehat{d}%
_{kl)}\right) \widehat{d}_{m}+ \\
\\
+12d_{m(ij}\widehat{d}_{kl)}-16\;\widehat{d}_{m(i}d_{jkl)}+ \\
\\
-12\;d_{p(ij}d_{kl)n}\;d_{mrs}\widehat{d}^{rp}\widehat{d}^{sn}
\end{array}
\right] .  \label{ETfinald-SKG}
\end{equation}
It is worth noticing that Eq. (\ref{ETfinald-SKG}) yields that the tensor
\begin{equation}
\widetilde{\overline{E}}_{\overline{m}ijkl,\xi =0}\equiv \nu ^{5/3}\overline{%
E}_{\overline{m}ijkl,\xi =0}  \label{E-tilde-csi=0}
\end{equation}
is independent on $\nu $, but it rather depends only on the \textit{%
``rescaled dilatons''} $\widehat{\lambda }^{i}$'s (recall definitions (\ref
{def-resc-dils})-(\ref{deff-2})):
\begin{equation}
\frac{\partial \widetilde{\overline{E}}_{\overline{m}ijkl,\xi =0}}{\partial
\nu }=0.
\end{equation}

By looking at Eq. (\ref{ETfinal}), it is easy to realize that the same does
not happen for $\xi \neq 0$: the non-vanishing of the quantum parameter $\xi
$ does not allow for an overall factorization of the dependence of $%
\overline{E}_{\overline{m}ijkl}$ on $\nu $ \textit{and/or} other (shifted
\textit{and/or} rescaled) variables. In other words, $\xi $ entangles the
dependence of $\overline{E}_{\overline{m}ijkl}$ on $\nu $ with the
dependence on $\widehat{\lambda }^{i}$'s, and thus the \textit{``}$\xi \neq
0 $\textit{\ analogue''} of $\widetilde{\overline{E}}_{\overline{m}ijkl,\xi
=0} $ (defined in (\ref{E-tilde-csi=0})) cannot be introduced. This fact is
related to the impossibility to uplift the quantum perturbatively corrected
SK geometry described by the prepotential (\ref{effe}) to $d=5$ space-time
dimensions. Indeed, as it is well known, in general only $d$-SK geometries
can be uplifted to $d=5$ (see \textit{e.g.} \cite{CFM1} and Refs. therein).

\section{\label{Sect-Curv}\textit{Sectional Curvature} at Critical Points}

In the present Section we reconsider the non-supersymmetric criticality
conditions for the effective BH potential $V_{BH}$ of an $\mathcal{N}=2$, $%
d=4$ Maxwell-Einstein supergravity coupled to a generic number $n_{V}$ of
Abelian vector supermultiplets. We will find that in both classes ($Z\neq 0$
and $Z=0$) of its non-BPS critical points, the critical value of $V_{BH}$
(and thus, through the Bekenstein-Hawking entropy-area formula, the
classical BH entropy) is proportional to the local value of the so-called
\textit{sectional curvature of matter charges}.

Within the present study, this general result then motivates the explicit
computation (carried out in the next Section in two different, but
equivalent, approaches) of the Riemann tensor, Ricci tensor and Ricci scalar
curvature for the SK geometries determined by the prepotential (\ref{effe}),
as well for generic $d$-SK geometry, obtained as the classical limit $\xi
\rightarrow 0$ of these former ones. This latter calculation extends to the
inclusion of the most general axion-shift-symmetric quantum perturbative
correction (see discussion in Introduction) the results on the curvature of
non-compact SK manifolds, found long time ago in \cite{CVP}.\medskip

Along the lines of the elaborations of \cite{Kallosh-review} (see also \cite
{CFMZ1}), we will now determine a \textit{``non-BPS }$Z=0$\textit{\
analogue''} of the \textit{``rule of three''} (\ref{rule-of-3}). Such a
\textit{``non-BPS }$Z=0$\textit{\ analogue''} is an hitherto unaddressed
issue in literature (for instance, not considered in the fairly general
treatment of \cite{TT1}, nor in \cite{Kallosh-review}). In order to derive
such a result, let us contract the constraints (\ref{SKG-constraints}) by $%
\overline{Z}^{i}Z^{\overline{j}}\overline{Z}^{k}Z^{\overline{l}}$, obtaining
\begin{equation}
R_{i\overline{j}k\overline{l}}\overline{Z}^{i}Z^{\overline{j}}\overline{Z}%
^{k}Z^{\overline{l}}=-2\left( Z_{i}\overline{Z}^{i}\right) ^{2}+C_{ikm}%
\overline{C}_{\overline{j}\overline{l}}^{~~~m}\overline{Z}^{i}Z^{\overline{j}%
}\overline{Z}^{k}Z^{\overline{l}}.  \label{5}
\end{equation}
Therefore, by recalling the non-BPS $Z=0$ criticality conditions for $V_{BH}$%
:
\begin{equation}
C_{ijk}\overline{Z}^{j}\overline{Z}^{k}=0,  \label{non-BPS-Z=0}
\end{equation}
as well as the definition of \textit{sectional curvature}\footnote{%
Notice that in general the Riemann tensor $R_{i\overline{j}k\overline{l}}$,
the Ricci tensor $R_{i\overline{j}}$, the Ricci scalar curvature $R$ and the
sectional curvature $\mathcal{R}$ itself all are real quantities.} (of the
\textit{matter charges}) (see \textit{e.g.} \cite{GRLS-1} for a recent use;
notice the different definition used here, consistent with the one adopted
in \cite{Kallosh-review}: see Eq. (3.1.1.2.11) therein)
\begin{equation}
\mathcal{R}\left( Z\right) \equiv R_{i\overline{j}k\overline{l}}\overline{Z}%
^{i}Z^{\overline{j}}\overline{Z}^{k}Z^{\overline{l}},  \label{def-sect-curv}
\end{equation}
it follows that at (\textit{``large''}) non-BPS $Z=0$ critical points of $%
V_{BH}$ it holds that:
\begin{equation}
\left( Z_{i}\overline{Z}^{i}\right) _{nBPS,Z=0}^{2}=\left[ g^{i\overline{j}%
}\left( \partial _{i}Z\right) \overline{\partial }_{\overline{j}}\overline{Z}%
\right] _{nBPS,Z=0}^{2}=-\frac{1}{2}\left. \mathcal{R}\left( Z\right)
\right| _{nBPS,Z=0}>0.  \label{sect-curv-non-BPS-Z=0}
\end{equation}
The result (\ref{sect-curv-non-BPS-Z=0}) holds for \textit{all} $\mathcal{N}%
=2$, $d=4$ \textit{ungauged} Maxwell-Einstein supergravities, not only for
the ones with symmetric scalar manifolds, and it implies that the \textit{%
sectional curvature of the matter charges} $\mathcal{R}\left( Z\right) $ to
be \textit{strictly negative} at non-BPS $Z=0$ critical points of $V_{BH}$.

Moreover, for \textit{symmetric} \texttt{(and actually also for homogeneous
non-symmetric...)} SK manifolds, recalling that along the non-BPS $Z=0$%
-supporting charge orbits the quartic invariant $\mathcal{I}_{4} $ is
positive, it further holds that (see \textit{e.g.} \cite{BFGM1} and \cite
{Kallosh-review})
\begin{equation}
\left[ g^{i\overline{j}}\left( \partial _{i}Z\right) \overline{\partial }_{%
\overline{j}}\overline{Z}\right] _{nBPS,Z=0}=\sqrt{-\frac{1}{2}\left.
\mathcal{R}\left( Z\right) \right| _{nBPS,Z=0}}=\sqrt{\mathcal{I}_{4}},
\end{equation}
thus yielding the relation
\begin{equation}
\left. \mathcal{R}\left( Z\right) \right| _{nBPS,Z=0}=-2\mathcal{I}_{4}<0.
\label{R-non-BPS-Z=0}
\end{equation}

Eq. (\ref{R-non-BPS-Z=0}) is to be contrasted with the analogue result
obtained in \cite{Kallosh-review} for (\textit{``large''}) non-BPS $Z\neq 0$
critical points of $V_{BH}$ in \textit{symmetric} SK geometries (see Eq.
(3.1.1.2.23), as well as Eq. (3.1.1.2.20), therein):
\begin{equation}
\left. \mathcal{R}\left( Z\right) \right| _{nBPS,Z\neq 0}=-6\left| Z\right|
_{nBPS,Z\neq 0}^{4}=\frac{3}{8}\mathcal{I}_{4}<0.  \label{6}
\end{equation}

Thus, \textit{at least} in \textit{symmetric} SK geometries, at various
classes of \textit{``large''} critical points of $V_{BH}$ the \textit{%
sectional curvature} \textit{of the matter charges} $\mathcal{R}\left(
Z\right) $ takes the following values:
\begin{equation}
\mathcal{R}\left( Z\right) =\left\{
\begin{array}{l}
\frac{1}{2}-BPS:0; \\
\\
nBPS,Z\neq 0:\frac{3}{8}\mathcal{I}_{4}<0; \\
\\
nBPS,Z=0:-2\mathcal{I}_{4}<0.
\end{array}
\right.  \label{RR}
\end{equation}
Correspondingly, through the celebrated Bekenstein-Hawking entropy-area
formula \cite{BH1} and its implementation through the \textit{Attractor
Mechanism} \cite{FGK}
\begin{equation}
S_{BH}=\pi \frac{A_{H}}{4}=\pi \left. V_{BH}\right| _{\partial V_{BH}=0},
\label{BH-e-a-formula}
\end{equation}
at (\textit{``large''}) non-BPS critical points of $V_{BH}$ in (\textit{at
least} symmetric) $\mathcal{N}=2$, $d=4$ \textit{ungauged} Maxwell-Einstein
supergravities, the value of the classical BH entropy is proportional to the
local value of $\mathcal{R}\left( Z\right) $ itself:
\begin{equation}
\frac{S_{BH}}{\pi }=\left\{
\begin{array}{l}
nBPS,Z\neq 0:2\sqrt{\frac{2}{3}}\sqrt{\left| \mathcal{R}\left( Z\right)
\right| }; \\
\\
nBPS,Z=0:\frac{1}{\sqrt{2}}\sqrt{\left| \mathcal{R}\left( Z\right) \right| }.
\end{array}
\right.  \label{S_BH}
\end{equation}
\medskip \medskip

Eqs. (\ref{RR}) (and consequently Eqs. (\ref{S_BH})) hold \textit{on-shell},
\textit{i.e.} at the various classes of critical points of $V_{BH}$.
Actually, they can be \textit{``unified''} into an \textit{off-shell} (%
\textit{i.e.} global) relation, involving $\mathcal{R}\left( Z\right) $
along with the true-vector (vanishing \textit{on-shell}) $\partial _{i}V_{BH}
$. In order to determine such a relation, let us evaluate the definition of
\textit{sectional curvature of matter charges} (\ref{def-sect-curv}) along
the constraints (\ref{SKG-constraints}), thus obtaining:
\begin{equation}
\mathcal{R}\left( Z\right) =-2\left( Z_{i}\overline{Z}^{i}\right) ^{2}+g^{n%
\overline{m}}C_{ikn}\overline{C}_{\overline{j}\overline{l}\overline{m}}%
\overline{Z}^{i}Z^{\overline{j}}\overline{Z}^{k}Z^{\overline{l}}.
\label{sect-curv-SKG}
\end{equation}
Now, by differentiating Eq. (\ref{VBH1}) and using the defining relations of
SK geometry (see \textit{e.g.} \cite{FerBig} and Refs. therein), one can
then write \cite{FGK}
\begin{equation}
D_{i}V_{BH}=\partial _{i}V_{BH}=2\overline{Z}Z_{i}+iC_{ijk}\overline{Z}^{j}%
\overline{Z}^{k}\Leftrightarrow C_{ijk}\overline{Z}^{j}\overline{Z}%
^{k}=-i\left( \partial _{i}V_{BH}-2\overline{Z}Z_{i}\right) .
\label{crit-cond--}
\end{equation}
By using Eq. (\ref{crit-cond--}), Eq. (\ref{sect-curv-SKG}) can thus be
recast in the following way:
\begin{eqnarray}
\mathcal{R}\left( Z\right)  &=&-2\left( Z_{i}\overline{Z}^{i}\right)
^{2}+g^{i\overline{j}}\left( \partial _{i}V_{BH}-2\overline{Z}Z_{i}\right)
\left( \overline{\partial }_{\overline{j}}V_{BH}-2Z\overline{Z}_{\overline{j}%
}\right) =  \notag \\
&=&2Z_{i}\overline{Z}^{i}\left( 2\left| Z\right| ^{2}-Z_{j}\overline{Z}%
^{j}\right) +  \notag \\
&&+g^{k\overline{l}}\left[ \left( \partial _{k}V_{BH}\right) \overline{%
\partial }_{\overline{l}}V_{BH}-2Z\left( \partial _{k}V_{BH}\right)
\overline{Z}_{\overline{l}}-2\overline{Z}\left( \overline{\partial }_{%
\overline{l}}V_{BH}\right) Z_{k}\right] .  \label{sect-curv-SKG-2}
\end{eqnarray}
Eq. (\ref{sect-curv-SKG-2}) is nothing but an equivalent rewriting of the
sectional curvature of atter charges in SK geometry, given by Eq. (\ref
{sect-curv-SKG}). By consistently using the criticality conditions of $V_{BH}
$ defining the various classes of (\textit{``large''}) critical points of $%
V_{BH}$ itself (namely: $\frac{1}{2}$-BPS - see Eq. (\ref{1/2-BPS-conds}) -,
non-BPS $Z=0$ - see Eq. (\ref{non-BPS-Z=0}) -, and non-BPS $Z\neq 0$ - see
Eq. (\ref{3-}) below), the three on-shell relations (\ref{RR}) are
obtained.\medskip

Aside, let us also notice that the constraints (\ref{SKG-constraints})
clearly yield a constrained expression for the Ricci tensor (and for the
Ricci scalar curvature) of a SK manifold, in which the partial (and
complete) contractions of the $C$-tensor with its complex conjugate play a
key role. Namely, Eq. (\ref{SKG-constraints}) respectively imply:
\begin{eqnarray}
R_{i\overline{j}} &\equiv &g^{k\overline{l}}R_{i\overline{l}k\overline{j}%
}=-\left( n_{V}+1\right) g_{i\overline{j}}+g^{k\overline{l}}g^{m\overline{n}%
}C_{imk}\overline{C}_{\overline{j}\overline{n}\overline{l}};\label{Ricci-SKG}
\\
R &\equiv &g^{i\overline{j}}g^{k\overline{l}}R_{i\overline{l}k\overline{j}%
}=g^{i\overline{j}}R_{i\overline{j}}=-\left( n_{V}+1\right) n_{V}+g^{i%
\overline{j}}g^{k\overline{l}}g^{m\overline{n}}C_{imk}\overline{C}_{%
\overline{j}\overline{n}\overline{l}}.\label{R-SKG}
\end{eqnarray}
From the discussion at the end of Subsect. \ref{First-Approach}, it will be
clear that the first terms in the right-hand sides of Eqs. (\ref{Ricci-SKG})
and (\ref{R-SKG}) are the constribution of the \textit{``quadratic sector''}
of the SK geometry (in which $C_{ijk}=0$, as a consequence of its very
definition (\ref{C}); notice that the contributions of such a \textit{%
``quadratic sector''} are missing in \textit{rigid} SK geometry, see \textit{%
e.g.} \cite{Freed} and \cite{Lu}).\medskip

A further elaboration for (\textit{``large''}) non-BPS $Z\neq 0$ critical
points of $V_{BH}$ can be performed by plugging the non-BPS $Z\neq 0$
criticality condition of $V_{BH}$ (see \textit{e.g.} \cite{Kallosh-review})
\begin{equation}
D_{i}\log Z=-\frac{i}{2}\frac{1}{\left| Z\right| ^{2}}C_{ijl}\overline{Z}^{j}%
\overline{Z}^{l}  \label{3-}
\end{equation}
into Eq. (\ref{5}), thus getting
\begin{gather}
\left. \mathcal{R}\left( Z\right) \right| _{nBPS,Z\neq 0}=\left[ 2Z_{i}%
\overline{Z}^{i}\left( 2\left| Z\right| ^{2}-Z_{i}\overline{Z}^{i}\right) %
\right] _{nBPS,Z\neq 0};  \label{7'} \\
\Updownarrow  \notag \\
\left( Z_{i}\overline{Z}^{i}\right) ^{2}-2g^{i\overline{j}}Z_{i}\overline{Z}%
^{i}\left| Z\right| ^{2}+\frac{1}{2}\mathcal{R}\left( Z\right) =0; \\
\Updownarrow  \notag \\
\left( Z_{i}\overline{Z}^{i}\right) _{\pm }=\left| Z\right| ^{2}\pm \sqrt{%
\left| Z\right| ^{4}-\frac{1}{2}\mathcal{R}\left( Z\right) };  \label{7''} \\
\Updownarrow  \notag \\
\left| Z\right| ^{2}=\frac{1}{4}\frac{\mathcal{R}\left( Z\right) }{Z_{i}%
\overline{Z}^{i}}+\frac{1}{2}Z_{i}\overline{Z}^{i},  \label{7}
\end{gather}
where the subscript ``$nBPS,Z\neq 0$'' has been suppressed for simplicity's
sake. Notice that, also when $\mathcal{R}\left( Z\right) >0$ satisfying $%
\left| Z\right| ^{4}-\frac{1}{2}\mathcal{R}\left( Z\right) \geqslant 0$,
only one brach of $Z_{i}\overline{Z}^{i}$ \textit{should} be consistent with
the fact that $Z_{i}\overline{Z}^{i}>0$.

Result (\ref{7'}), holding for all $\mathcal{N}=2$, $d=4$ \textit{ungauged}
Maxwell-Einstein supergravities, not only for the ones with symmetric scalar
manifolds, consistently reduces to Eq. (\ref{6}) when the \textit{``rule of
three''} (\ref{rule-of-3})holds, as it is the case for symmetric SK
manifolds (see discussion above).

By recalling Eqs. (\ref{V-1}) and (\ref{Delta}), Eq. (\ref{7}) thus implies
that
\begin{equation}
\left. \mathcal{R}\left( Z\right) \right| _{non-BPS,Z\neq 0}=-\left[ 2\left(
3+\frac{\Delta }{\left| Z\right| ^{2}}\right) \left( 1+\frac{\Delta }{\left|
Z\right| ^{2}}\right) \left| Z\right| ^{4}\right] _{non-BPS,Z\neq 0},
\label{Pp-1}
\end{equation}
or, more explicitly (evaluation at \textit{``large''} non-BPS $Z\neq 0$
critical points of $V_{BH}$ understood)
\begin{equation}
\mathcal{R}\left( Z\right) =-\frac{9}{8}\frac{\left| Z\right| ^{4}}{\left(
\overline{N_{3}}\left( Z\right) \right) ^{2}\left| Z\right| ^{4}}\left[ 4%
\overline{N_{3}}\left( Z\right) \left| Z\right| ^{2}-E\left( Z,\overline{Z}%
\right) \right] \left[ \frac{4}{3}\overline{N_{3}}\left( Z\right) \left|
Z\right| ^{2}-E\left( Z,\overline{Z}\right) \right] .  \label{Pp-2}
\end{equation}
where
\begin{eqnarray}
\overline{N_{3}}\left( Z\right) &\equiv &\overline{C}_{\overline{i}\overline{%
j}\overline{k}}Z^{\overline{i}}Z^{\overline{j}}Z^{\overline{k}}; \\
E\left( Z,\overline{Z}\right) &\equiv &E_{i\overline{j}\overline{k}\overline{%
l}\overline{m}}\overline{Z}^{i}Z^{\overline{j}}Z^{\overline{k}}Z^{\overline{l%
}}Z^{\overline{m}}.
\end{eqnarray}
Results (\ref{Pp-1}) and (\ref{Pp-2}) relate $\mathcal{R}\left( Z\right) $, $%
\overline{N_{3}}\left( Z\right) $ and $E\left( Z,\overline{Z}\right) $ at
\textit{``large''} non-BPS $Z\neq 0$ critical points of $V_{BH}$ in generic $%
\mathcal{N}=2$, $d=4$ \textit{ungauged} Maxwell-Einstein supergravities, and
they consistently reduce to Eq. (\ref{6}) (\textit{at least}) for \textit{%
symmetric} SK manifolds. They are consistent with the treatment performed in
\cite{ADFT-review,Kallosh-review,CFMZ1}, see e.g. Eq. (3.1.1.2.17) of \cite
{CFMZ1}, here reported for ease of comparison (evaluation at non-BPS $Z\neq
0 $ critical points of $V_{BH}$ understood):
\begin{equation}
\frac{3}{4}\frac{1}{\left| Z\right| ^{2}}\frac{E\left( Z,\overline{Z}\right)
}{\overline{N_{3}}\left( Z\right) }-1=\frac{\mathcal{R}\left( Z\right) }{%
2\left| Z\right| ^{2}Z_{i}\overline{Z}^{i}}=2\frac{\mathcal{R}\left(
Z\right) }{C_{ikn}\overline{C}_{\overline{r}\overline{s}}^{i}\overline{Z}^{k}%
\overline{Z}^{n}Z^{\overline{r}}Z^{\overline{s}}}.  \label{UCLA11}
\end{equation}
Furthermore, through the definition (\ref{Delta}), Eqs. (\ref{Pp-1}) and (%
\ref{Pp-2}) are implied also by Eq. (\ref{7}). Notice that, while $\mathcal{R%
}\left( Z\right) $ is a real quantity, $E_{i\overline{j}\overline{k}%
\overline{l}\overline{m}}$, $E\left( Z,\overline{Z}\right) $ and $\Delta $
are generally complex. But, (at least) at non-BPS $Z\neq 0$ critical points,
$\Delta $, or equivalently the ratio $\frac{E\left( Z,\overline{Z}\right) }{%
\overline{N_{3}}\left( Z\right) }$, becomes real (consistent with Eq. (\ref
{Delta}); see also Eq. (276) of \cite{ADFT-review} and Eq. (5.17) of \cite
{CFMZ1}).

\section{\label{Riemann-Tensor}Riemann Tensor}

The new results obtained in previous Section call for an explicit
computation of the Riemann tensor, Ricci tensor and Ricci scalar curvature
for the SK geometries determined by the prepotential (\ref{effe}), as well
for generic $d$-SK geometry, obtained as the classical limit $\xi
\rightarrow 0$ of these former ones. We will do this in the present Section,
carrying out the calculation in two different, but (proved to be)
equivalent, ways.

\subsection{\label{First-Approach}First Approach}

The first approach conceives SK geometry as a particular K\"{a}hler
geometry, and therefore one starts with nothing but the standard formula of
Riemann tensor:
\begin{equation}
R_{i\overline{j}k\overline{l}}\equiv g^{m\overline{n}}\left( \overline{%
\partial }_{\overline{l}}\overline{\partial }_{\overline{j}}\partial
_{m}K\right) \partial _{i}\overline{\partial }_{\overline{n}}\partial _{k}K-%
\overline{\partial }_{\overline{l}}\partial _{i}\overline{\partial }_{%
\overline{j}}\partial _{k}K.  \label{Riemann-Kahler}
\end{equation}

After long but straightforward algebra (detailed in Appendix C), the Riemann
tensor of the SK geometry determined by the prepotential (\ref{effe}) is
computed as
\begin{eqnarray}
R_{i\overline{j}k\overline{l}} &=&R_{ijkl}=-\frac{\nu ^{2/3}}{32\left( \nu -%
\frac{\xi }{2}\right) ^{2}}\cdot  \notag \\
&&\cdot \left\{
\begin{array}{l}
-\frac{\left( \nu -\frac{\xi }{2}\right) }{\nu +\xi }\widehat{d}_{ik}%
\widehat{d}_{jl}+2\widehat{d}_{ij}\widehat{d}_{kl}+2\widehat{d}_{il}\widehat{%
d}_{jk}+ \\
\\
+\frac{\nu ^{2}}{4\left( \nu -\frac{\xi }{2}\right) ^{2}}\widehat{d}_{i}%
\widehat{d}_{j}\widehat{d}_{k}\widehat{d}_{l}+ \\
\\
-\frac{\nu }{2\left( \nu -\frac{\xi }{2}\right) }\left( \widehat{d}_{ij}%
\widehat{d}_{k}\widehat{d}_{l}+\widehat{d}_{jk}\widehat{d}_{i}\widehat{d}%
_{l}+\widehat{d}_{il}\widehat{d}_{j}\widehat{d}_{k}+\widehat{d}_{kl}\widehat{%
d}_{i}\widehat{d}_{j}\right) + \\
\\
+2\frac{\left( \nu -\frac{\xi }{2}\right) }{\nu }d_{ikn}d_{jlm}\widehat{d}%
^{mn}
\end{array}
\right\} .  \label{Riemann-csi}
\end{eqnarray}
In the classical limit ($\xi \rightarrow 0$), the expression of the Riemann
tensor in a generic $d$-SK geometry is easily obtained:
\begin{eqnarray}
R_{i\overline{j}k\overline{l},\xi =0} &=&R_{ijkl,\xi =0}=-\frac{1}{32}\nu
^{-4/3}\cdot  \notag \\
&&\cdot \left\{
\begin{array}{l}
-\widehat{d}_{ik}\widehat{d}_{jl}+2\widehat{d}_{ij}\widehat{d}_{kl}+2%
\widehat{d}_{il}\widehat{d}_{jk}+\frac{1}{4}\widehat{d}_{i}\widehat{d}_{j}%
\widehat{d}_{k}\widehat{d}_{l}+ \\
\\
-\frac{1}{2}\left( \widehat{d}_{ij}\widehat{d}_{k}\widehat{d}_{l}+\widehat{d}%
_{jk}\widehat{d}_{i}\widehat{d}_{l}+\widehat{d}_{il}\widehat{d}_{j}\widehat{d%
}_{k}+\widehat{d}_{kl}\widehat{d}_{i}\widehat{d}_{j}\right) +2d_{ikn}d_{jlm}%
\widehat{d}^{mn}
\end{array}
\right\} .  \label{Riemann-csi=0}
\end{eqnarray}
Notice that both Eqs. (\ref{Riemann-csi}) and (\ref{Riemann-csi=0}) have all
the symmetry properties suitable to the Riemann tensor.

Consequently, the Ricci tensor and Ricci curvature scalar can respectively
be computed as follows (recall $n_{V}$ denotes the number of Abelian vector
multiplets coupled to gravity multiplet, or equivalently the complex
dimension of the considered SK manifold):
\begin{eqnarray}
R_{i\overline{j}} &\equiv &g^{k\overline{l}}R_{i\overline{l}k\overline{j}}=-%
\frac{1}{16}\frac{\nu ^{4/3}}{\left( \nu -\frac{\xi }{2}\right) ^{2}}\left[
\begin{array}{l}
\frac{\left[ n_{V}\nu ^{3}+\frac{3}{2}\left( n_{V}+2\right) \nu ^{2}\xi -%
\frac{3}{4}\nu \xi ^{2}-\frac{1}{8}\left( 4n_{V}+3\right) \xi ^{3}\right] }{%
\left( \nu +\xi \right) ^{2}\left( \nu -\frac{\xi }{2}\right) }\widehat{d}%
_{i}\widehat{d}_{j}+ \\
\\
-\frac{\left[ 4n_{V}\nu ^{2}+2\left( n_{V}+3\right) \nu \xi -\left(
2n_{V}+3\right) \xi ^{2}\right] }{\nu \left( \nu +\xi \right) }\widehat{d}%
_{ij}+ \\
\\
-4\frac{\left( \nu -\frac{\xi }{2}\right) ^{2}}{\nu ^{2}}d_{ikn}d_{jlm}%
\widehat{d}^{kl}\widehat{d}^{mn}
\end{array}
\right] =  \notag \\
&&~  \notag \\
&&~  \notag \\
&=&-\frac{1}{16}\frac{\left[ n_{V}\nu ^{3}+\frac{3}{2}\left( n_{V}+2\right)
\nu ^{2}\xi -\frac{27}{36}\nu \xi ^{2}-\frac{1}{8}\left( 4n_{V}+3\right) \xi
^{3}\right] }{\left( \nu -\frac{\xi }{2}\right) ^{3}\left( \nu +\xi \right)
^{2}}\nu ^{4/3}\widehat{d}_{i}\widehat{d}_{j}+  \notag \\
&&  \notag \\
&&+\frac{1}{16}\frac{\left[ 4n_{V}\nu ^{2}+2\left( n_{V}+3\right) \nu \xi
-\left( 2n_{V}+3\right) \xi ^{2}\right] }{\left( \nu -\frac{\xi }{2}\right)
^{2}\left( \nu +\xi \right) }\nu ^{1/3}\widehat{d}_{ij}+  \notag \\
&&  \notag \\
&&+\frac{1}{4}\nu ^{-2/3}d_{ikn}d_{jlm}\widehat{d}^{kl}\widehat{d}^{mn}
\notag \\
&=&R_{ij}  \label{Ricci-csi} \\
&&  \notag \\
R &\equiv &g^{i\overline{j}}R_{i\overline{j}}=-n_{V}\left( n_{V}+1\right) -%
\frac{9}{2}\frac{\left( \nu -\frac{\xi }{2}\right) \nu }{\left( \nu +\xi
\right) ^{3}}\xi +\frac{3}{2}n_{V}\frac{\left( \nu -\frac{\xi }{2}\right) }{%
\nu +\xi }-\frac{\left( \nu -\frac{\xi }{2}\right) }{\nu }d_{ikn}d_{jlm}%
\widehat{d}^{jk}\widehat{d}^{mn}\widehat{d}^{il}.  \notag \\
&&  \label{R-csi}
\end{eqnarray}
Thence, in the classical limit ($\xi \rightarrow 0$), the expression of the
Ricci tensor and Ricci scalar curvature in a generic $d$-SK geometry is
easily obtained, respectively:
\begin{eqnarray}
R_{i\overline{j},\xi =0} &\equiv &\breve{g}^{kl}R_{ilkj,\xi =0}=-\frac{1}{16}%
\nu ^{-2/3}\left( n_{V}\widehat{d}_{i}\widehat{d}_{j}-4n_{V}\widehat{d}%
_{ij}-4d_{ikn}d_{jlm}\widehat{d}^{kl}\widehat{d}^{mn}\right) =R_{ij,\xi =0};
\label{Ricci-csi=0} \\
&&  \notag \\
R_{\xi =0} &\equiv &\breve{g}^{ij}R_{ij,\xi =0}=-n_{V}\left( n_{V}+1\right) +%
\frac{3}{2}n_{V}-d_{ikn}d_{jlm}\widehat{d}^{jk}\widehat{d}^{mn}\widehat{d}%
^{il}=  \notag \\
&=&-n_{V}^{2}+\frac{n_{V}}{2}-d_{ikn}d_{jlm}\widehat{d}^{jk}\widehat{d}^{mn}%
\widehat{d}^{il}.  \label{R-csi=0}
\end{eqnarray}
Let us notice that both Eqs. (\ref{Ricci-csi}) and (\ref{Ricci-csi=0}) have
the symmetry properties suitable for Ricci tensor.

As pointed out at the end of Sect. \ref{E-Tensor}, the symmetricity
conditions (\ref{DR=0})-(\ref{symm}) cannot be satisfied for prepotential (%
\ref{effe}) with $\xi \neq 0$. As it is well known, all symmetric spaces are
Einstein spaces (see \textit{e.g.} \cite{Helgason}, and \cite{LA08-Proc} for
a comprehensive list of Refs.), \textit{i.e.} with a Ricci tensor satisfying
\begin{equation}
\exists \Lambda \in \mathbb{R}:R_{i\overline{j}}=\Lambda g_{i\overline{j}%
}\Rightarrow R=n_{V}\Lambda ,  \label{Einstein}
\end{equation}
and then with a constant Ricci scalar curvature, whose sign is the one of
the real constant $\Lambda $ itself. However, the opposite does not generally
hold true: not all Einstein spaces are symmetric. Thus, it is reasonable to
ask whether the considered quantum SK geometries determined by prepotential (%
\ref{effe}) can be Einstein. By recalling Eq. (\ref{covmetric}) and using
Eq. (\ref{Ricci-csi}), the condition for such geometries to be Einstein can
be written as follows:
\begin{equation}
\frac{1}{4}\frac{\nu }{\left( \nu -\frac{\xi }{2}\right) }\left[
\begin{array}{l}
\frac{\left[ n_{V}\nu ^{3}+\frac{3}{2}\left( n_{V}+2\right) \nu ^{2}\xi -%
\frac{3}{4}\nu \xi ^{2}-\frac{1}{8}\left( 4n_{V}+3\right) \xi ^{3}\right] }{%
\left( \nu +\xi \right) ^{2}\left( \nu -\frac{\xi }{2}\right) }\widehat{d}%
_{i}\widehat{d}_{j}+ \\
\\
-\frac{\left[ 4n_{V}\nu ^{2}+2\left( n_{V}+3\right) \nu \xi -\left(
2n_{V}+3\right) \xi ^{2}\right] }{\nu \left( \nu +\xi \right) }\widehat{d}%
_{ij}+ \\
\\
-4\frac{\left( \nu -\frac{\xi }{2}\right) ^{2}}{\nu ^{2}}d_{ikn}d_{jlm}%
\widehat{d}^{kl}\widehat{d}^{mn}
\end{array}
\right] =\Lambda \left[ \widehat{d}_{ij}-\frac{1}{4}\frac{\nu }{(\nu -\frac{1%
}{2}\xi )}\widehat{d}_{i}\widehat{d}_{j}\right] ,  \label{Einstein-csi}
\end{equation}
and it seems to us that such an Eq. does not admit solutions for any value
of the real constants $\xi $ and $\Lambda $.

The situation is pretty different for the classical limit ($\xi \rightarrow
0 $), determining the so-called $d$-SK geometries (described by prepotential
(\ref{effe}) with $\xi =0$). For such geometries, by recalling Eq. (\ref{3})
and using Eq. (\ref{Ricci-csi=0}), the condition to be Einstein reads
\begin{equation}
\frac{1}{4}\left( n_{V}\widehat{d}_{i}\widehat{d}_{j}-4n_{V}\widehat{d}%
_{ij}-4d_{ikn}d_{jlm}\widehat{d}^{kl}\widehat{d}^{mn}\right) =\Lambda \left(
\widehat{d}_{ij}-\frac{\widehat{d}_{i}\widehat{d}_{j}}{4}\right) .
\label{Einstein-csi=0}
\end{equation}
As found in \cite{CVP} (see also \cite{dWVP}), a (proper) subset of
solutions to Eq. (\ref{Einstein-csi=0}) is given by the symmetric $d$-SK
geometries, satisfying the conditions of symmetricity (\ref{DR=0})-(\ref
{symm}). (\textit{At least}) in such geometries, the $d$-tensor satisfies
the following relation (\cite{CVP,BFGM1,CFM1}; see also the treatment given
in \cite{Kallosh-review}, and Refs. therein):
\begin{equation}
d_{p(kl}d_{ij)n}a^{pr}a^{ns}a^{mq}d_{rsq}=\frac{4}{3}\delta
_{(k}^{m}d_{lij)},  \label{rel-1}
\end{equation}
which is a consequence of Eq. (\ref{symm}), and in fact can be further
elaborated by using the second relation (involving the Riemann tensor) in
Eq. (\ref{symm}) itself. In Eq. (\ref{rel-1}) $a^{ij}$ is a sort of \textit{%
``rescaled''} metric tensor, defined as (recall Eq. (\ref{4}); see \textit{%
e.g.} \cite{CFM1} for further elucidation of $d=5$ origin of such a
quantity):
\begin{equation}
a^{ij}\equiv \frac{1}{4}\nu ^{-2/3}\breve{g}^{ij}=\frac{1}{2}\left( \widehat{%
\lambda }^{i}\widehat{\lambda }^{j}-2\widehat{d}^{ij}\right) .
\end{equation}

Let us also notice that, from Eq. (\ref{Einstein}) the constancy of the
Ricci scalar curvature is necessary but not sufficient condition for
Einstein, and in turn for symmetric, spaces. In other words, it holds:
\begin{equation}
symmetric\overset{\nLeftarrow }{\Longrightarrow }Einstein\overset{%
\nLeftarrow }{\Longrightarrow }constant~R.
\end{equation}
Eq. (\ref{R-csi}) yields the condition ($\Omega \in \mathbb{R}$)
\begin{equation}
-n_{V}\left( n_{V}+1\right) -\frac{9}{2}\frac{\left( \nu -\frac{\xi }{2}%
\right) \nu }{\left( \nu +\xi \right) ^{3}}\xi +\frac{3}{2}n_{V}\frac{\left(
\nu -\frac{\xi }{2}\right) }{\nu +\xi }-\frac{\left( \nu -\frac{\xi }{2}%
\right) }{\nu }d_{ikn}d_{jlm}\widehat{d}^{jk}\widehat{d}^{mn}\widehat{d}%
^{il}=\Omega ,  \label{R-const-csi}
\end{equation}
and it seems that it is not possible to have $R$ constant for SK geometries
determined by (\ref{effe}) with $\xi \neq 0$. On the other hand, Eq. (\ref
{R-csi=0}) yields the condition
\begin{equation}
R_{\xi =0}=-n_{V}^{2}+\frac{n_{V}}{2}-d_{ikn}d_{jlm}\widehat{d}^{jk}\widehat{%
d}^{mn}\widehat{d}^{il}=\Omega .  \label{R-const-csi=0}
\end{equation}
As pointed out above, a (proper) set of solutions to condition (\ref
{R-const-csi=0}) is given by the symmetric $d$-SK geometries. As for all
Einstein spaces, for symmetric $d$-SK spaces it holds that
\begin{equation}
\Omega =\Lambda n_{V}.  \label{Omega-Einstein}
\end{equation}
The results of \cite{CVP} yields $\Lambda =-\frac{2}{3}n_{V}$ for the four
irreducible symmetric $d$-SK geometries (which are nothing but the \textit{%
``magic''} ones) and $\Lambda =-\frac{\left( n_{V}^{2}-2n_{V}+3\right) }{%
n_{V}}$ for the reducible sequence $\frac{SU(1,1)}{U(1)}\times \ \frac{%
SO\left( 2,n_{V}-1\right) }{SO\left( 2\right) \times SO\left( n_{V}-1\right)
}$ (and $\Lambda =-\left( n_{V}+1\right) $ for the \textit{minimal coupling}
$\mathbb{C}\mathbb{P}^{n_{V}}$ sequence, whose prepotential is however
\textit{quadratic}).\medskip

Analogously to the comment made at the and of Sect. \ref{E-Tensor}, it is
here worth noticing that Eqs. (\ref{Riemann-csi=0}), (\ref{Ricci-csi=0}) and
(\ref{R-csi=0}) respectively yield that the quantities
\begin{eqnarray}
\widetilde{R}_{i\overline{j}k\overline{l},\xi =0} &\equiv &\nu ^{4/3}R_{i%
\overline{j}k\overline{l},\xi =0};  \label{Riemann-tilde-csi=0} \\
\widetilde{R}_{i\overline{j},\xi =0} &\equiv &\nu ^{2/3}R_{i\overline{j},\xi
=0};  \label{Ricci-tilde-csi=0} \\
&&R_{\xi =0};
\end{eqnarray}
are independent on $\nu $, but they rather depend only on the \textit{%
``rescaled dilatons''} $\widehat{\lambda }^{i}$'s (recall definitions (\ref
{def-resc-dils})-(\ref{deff-2})):
\begin{eqnarray}
\frac{\partial \widetilde{R}_{i\overline{j}k\overline{l},\xi =0}}{\partial
\nu } &=&0; \\
\frac{\partial \widetilde{R}_{i\overline{j},\xi =0}}{\partial \nu } &=&0; \\
\frac{\partial R_{\xi =0}}{\partial \nu } &=&0.
\end{eqnarray}

By looking at Eqs. (\ref{Riemann-csi}), (\ref{Ricci-csi}) and (\ref{R-csi}),
it is easy to realize that the same does not happen for $\xi \neq 0$: the
non-vanishing of the quantum parameter $\xi $ does not allow for an overall
factorization of the dependence of $R_{i\overline{j}k\overline{l}}$, $R_{i%
\overline{j},\xi =0}$ and $R$ on $\nu $ \textit{and/or} other (shifted
\textit{and/or} rescaled) variables. In other words, $\xi $ entangles the
dependence of $R_{i\overline{j}k\overline{l}}$, $R_{i\overline{j},\xi =0}$
and $R$ on $\nu $ with the dependence on $\widehat{\lambda }^{i}$'s, and
thus the \textit{``}$\xi \neq 0$\textit{\ analogues''} of $\widetilde{R}_{i%
\overline{j}k\overline{l},\xi =0}$ and $\widetilde{R}_{i\overline{j},\xi =0}$
(respectively defined in (\ref{Riemann-tilde-csi=0}) and (\ref
{Ricci-tilde-csi=0})) cannot be introduced. As already pointed out at the
end of Sect. \ref{E-Tensor}, this fact is related to the impossibility to
uplift the quantum perturbatively corrected SK geometry described by the
prepotential (\ref{effe}) to $d=5$ space-time dimensions. Indeed, as it is
well known, in general only $d$-SK geometries can be uplifted to $d=5$ (see
\textit{e.g.} \cite{CFM1} and Refs. therein).

\subsection{\label{Second-Approach}Second Approach}

The second approach is actually the one considered in \cite{CVP}: the
constraints (\ref{SKG-constraints}), characterizing, among others, a
K\"{a}hler geometry to be \textit{special}, are exploited in order to
compute the Riemann tensor itself, yielding the same results given by Eq. (%
\ref{Riemann-csi}) and (\ref{Riemann-csi=0}), respectively for the
prepotential (\ref{effe}) and its classical limit $\xi \rightarrow 0$ ($d$%
-SK geometry). The same can explicitly be proved to hold for the Ricci
tensor (\ref{Ricci-csi}) and the Ricci scalar (\ref{R-csi}), and for their
respective classical limits (\ref{Ricci-csi=0}) and (\ref{R-csi=0}).

Thus, the approaches respectively based on (\ref{Riemann-Kahler}) and (\ref
{SKG-constraints}) have been proved to be equivalent, by explicitly
computing the expressions of the Riemann tensor $R_{i\overline{j}k\overline{l%
}}$, of Ricci tensor $R_{i\overline{j}}$ and of Ricci scalar curvature $R$
of a SK geometry of arbitrary complex dimension $n_{V}$ and determined by
the holomorphic prepotential (\ref{effe}) (also considering the
corresponding limit of $d$-SK geometry, obtained by letting the quantum
parameter $\xi \rightarrow 0$). As previously mentioned, by including in the
prepotential the most general quantum perturbative correction consistent
with the Peccei-Quinn axion-shift symmetry \cite{Peccei-Quinn} (see
discussion in the Introduction), the results and considerations of Sect. \ref
{Riemann-Tensor} are an extension of the findings of \cite{CVP} to the quantum
perturbative regime.

\section{\label{Conclusion}Conclusion}

It is clear that the present investigation (completing, extending and
generalizing the work of \cite{BFMS1} and \cite{BFMS2}) does not conclude
the study of quantum (perturbative) SK geometries. Only some venues have
been considered in the vast realm of quantum geometries of the moduli spaces
of superstring theories. Many issues still deserve a deeper understanding
and call for a thorough analysis, and we leave them for further future
study. Below, we list only some of the most appealing ones (to us).

\begin{enumerate}
\item  It would be interesting to determine the extent of validity of the
so-called \textit{``rule of three''} (\ref{rule-of-3}), which is nothing but
the sum rule determining the value of $V_{BH}$ at its non-BPS $Z\neq 0$
critical points. While its \textit{``non-BPS }$Z=0$\textit{\ analogue''} (%
\ref{sect-curv-non-BPS-Z=0}) has general validity, (\ref{rule-of-3}) does
not hold in general. Firstly noticed in \cite{TT1}, the \textit{``rule of
three''} (\ref{rule-of-3}) has been proved to hold in \textit{symmetric} SK
geometries \cite{BFGM1}, in (\textit{at least} some of the) homogeneous
\textit{non-symmetric} $d$-SK geometries \cite{DFT-hom-non-symm} (and in $%
\mathcal{N}>2$-extended supergravities admitting non-supersymmetric
attractors with non-vanishing central charge matrix \cite{FK-N=8,ADFT-review}%
). The most general results for $d$-SK geometries currently available are
given in \cite{TT1}, but they are depending on the particular considered BH
charge configurations; thus, it would be nice to see whether the \textit{%
``rule of three''} (\ref{rule-of-3}) still holds in a generic BH charge
configuration. On the other hand, since the condition (\ref{rule-of-3-1}) of
validity of the \textit{``rule of three''} does not imply symmetricity (nor
homogeneity), it would be nice to see if and how the \textit{``rule of
three''} works in the quantum corrected SK geometries (\ref{effe}).

\item  In the present paper we explained the peculiarity of the $D0-D6$
configuration in presence of the most general axion-shift-symmetric quantum
perturbative parameter $\xi $. The $D0-D6$ configuration turns out to be the
somewhat \textit{``minimal''} configuration which does \textit{not} support
axion-free critical points of $V_{BH}$. But we did not yet completely
explain the results of the investigation of \cite{BFMS2}. Namely, we did not
explain why the classical non-BPS $Z\neq 0$ \textit{``flat''} direction of $%
V_{BH}$ of the $st^{2}$ model gets non-renormalized (despite acquiring a
non-vanishing axion) when switching $\xi $ on. We leave the investigation of
this issue (within $d$-SKG geometries of arbitrary complex dimension $n_{V}$%
) for future study.

\item  An issue concerning both $d$-SK geometries and their quantum
corrected counterparts (\ref{effe}) is the generality of the \textit{%
axion-free} solutions (\textit{if any}) to the Attractor Eqs.. As found in
\cite{CFM1}, the axion-free-supporting BH charge configurations in $d$-SK
geometries are the \textit{electric} ($D2-D6$), \textit{magnetic} ($D0-D4$)
and $D0-D6$ ones, whereas in the present work we obtained that for SK
geometries determined by the prepotential (\ref{effe}) only \textit{electric}
and \textit{magnetic} configurations support purely imaginary critical points of $V_{BH}$.
It would be interesting to analyze the degree of generality of \textit{%
axion-free} solutions (in a model-independent fashion, if possible) in these
frameworks.

\item  Concerning $d$-SKG geometries, the expression of the $\frac{1}{2}$%
-BPS attractors is known in the most explicit form possible \cite{Shmakova},
and (going beyond symmetric cases) there are various explicit (but
charge-dependent) results for non-BPS $Z\neq 0$ critical points of $V_{BH}$
(see \textit{e.g.} \cite{TT1}). On the other hand, there are currently no
general results on the explicit form of non-BPS $Z=0$ critical points of $%
V_{BH}$ within the same SK geometry. Thus, it would be interesting to
determine such expression and use it to elaborate the \textit{``non-BPS }$%
Z=0 $\textit{\ analogue''} (\ref{sect-curv-non-BPS-Z=0}) (obtained in the
present paper) of the \textit{``rule of three''} (\ref{rule-of-3}).

\item  Still very little is known on the explicit expression of the critical
points of the quantum perturbatively corrected BH potential $V_{BH}$ given
by Eq. (\ref{V_BH-d-SKG-lambda}). The complete analysis of $\frac{1}{2}$-BPS
critical points (beyond the axion-free results of \cite{CQ-N=2-BHs}; see the
end of Sect. \ref{Z-and-DZ}) should be based on the implementation of $\frac{%
1}{2}$-BPS conditions (\ref{1/2-BPS-conds}) through the formula (\ref{DZ}).
More interestingly, the non-BPS ($Z\neq 0$ and $Z=0$) critical points of (%
\ref{V_BH-d-SKG-lambda}) still need to be completely determined and studied.

\item  The phenomena of \textit{``splitting''} of attractors \cite{BFMS1},
\textit{``transmutation''} of attractors \cite{BFMS1}, and \textit{%
``lifting''} of \textit{moduli spaces} of attractors \cite{BFMS2}, even if
explicitly found by studying models with only one or two complex scalar
field(s), are likely to characterize the quantum perturbatively corrected SK
geometry (\ref{effe}) for an arbitrary complex dimension $n_{V}$. Thus, it
would be worth studying more in depth such phenomena, eventually relating
them with the presence of particular symmetry groups acting in transitive or
non-transitive way on the (generally non-homogeneous) scalar manifold.

\item  By extending the results obtained in \cite{BFMS2} (at least in the
\textit{magnetic} and \textit{electric} configurations) to the presence of
more than one \textit{``flat''} direction, and including the effects of
non-perturbative corrections (see \textit{e.g.} \cite
{CDLOGP1,CQ-N=2-BHs,Behrndt-3,Cardoso-2}), one would lead to conjecture
that \textit{only a (very) few classical attractors do remain attractors in
strict sense at the quantum level}. Consequently, \textit{at the quantum}
(perturbative \textit{and} non-perturbative) \textit{level the set of actual
extremal BH attractors should be strongly constrained and reduced}. As
already noticed in the Conclusion of \cite{BFMS2} itself, in $\mathcal{N}=8$%
, $d=4$ supergravity the (\textit{``large''}) $\frac{1}{8}$-BPS and non-BPS
BHs critical points of $V_{BH,\mathcal{N}=8}$ exhibit $40$ and $42$ \textit{%
``flat''} directions, respectively \cite{ADF-Duality-d=4,Ferrara-Marrani-1}.
Within the possibility of $\mathcal{N}=8$ supergravity to be a finite theory
of quantum gravity (see \textit{e.g.} \cite{Bern} and \cite{K-N=8}, and
Refs. therein), it would be interesting to understand whether these \textit{%
``flat''} directions may be removed at all by perturbative \textit{and/or}
non-perturbative quantum effects.

\end{enumerate}

\section*{Acknowledgments}

We would like to thank S. Ferrara, E. Orazi, A. Shcherbakov, A. Yeranyan for
enlightening discussions.

A. M. would like to thank the \textit{Center for Theoretical Physics} (CTP)
of the University of California, Berkeley, CA USA, where part of this work
was done, for kind hospitality and stimulating environment. Furthermore, A.
M. would like to thank Ms. Hanna Hacham for peaceful and inspiring
hospitality in Palo Alto, CA USA.

R.R. would like to thank INFN - Frascati National Laboratories for
kind hospitality and support.

This work is supported in part by the ERC Advanced
Grant no. 226455, \textit{``Supersymmetry, Quantum Gravity and Gauge Fields''%
} (\textit{SUPERFIELDS}).

The work of S. B. ~has been supported in part by the grant INTAS-05-7928.

The work of A. M. has been supported by an INFN visiting Theoretical
Fellowship at SITP, Stanford University, Stanford, CA, USA.

The work of R. R. has been supported in part by Dipartimento di Scienze
Fisiche of Federico II University and INFN- Section of Napoli. \appendix

\section{Details of Computation of $V_{BH}$}

The symplectic-covariant holomorphic sections determined by the prepotential
(\ref{prepot}) read
\begin{equation}
F_{\Lambda }(X;\xi )=\frac{\partial F(X;\xi )}{\partial X^{\Lambda }}%
:\left\{
\begin{array}{l}
\Lambda =0\;\;:F_{0}(X;\xi )=-\frac{1}{3!}d_{ijk}\frac{X^{i}X^{j}X^{k}}{%
(X^{0})^{2}}+2i\xi X^{0}; \\
\\
\Lambda =i\;\;:F_{i}(X)=\frac{1}{2}d_{ijk}\frac{X^{j}X^{k}}{X^{0}},
\end{array}
\right.
\end{equation}
thus yielding the following check of homogeneity of degree $2$ of $F(X;\xi )$
in $X^{\Lambda }$'s:
\begin{equation}
F_{\Lambda }(X;\xi )X^{\Lambda }=F_{0}(X;\xi )X^{0}+F_{i}(X)X^{i}=2F(X;\xi ).
\end{equation}

Let us now move to evaluate the various components of the symmetric matrix:
\begin{equation}
\mathcal{F}_{\Lambda \Sigma }(X;\xi )=\frac{\partial F_{\Lambda }(X;\xi )}{%
\partial X^{\Sigma }}=\frac{\partial F_{\Sigma }(X;\xi )}{\partial
X^{\Lambda }}=\frac{\partial ^{2}F(X;\xi )}{\partial X^{\Lambda }\partial
X^{\Sigma }}=\frac{\partial ^{2}F(X;\xi )}{\partial X^{(\Lambda }\partial
X^{\Sigma )}}.
\end{equation}
They read:
\begin{eqnarray}
&(\Lambda ,\Sigma )=&(0,0)\;\;:\mathcal{F}_{00}(X;\xi )=\frac{1}{3}d_{ijk}%
\frac{X^{i}X^{j}X^{k}}{(X^{0})^{3}}+2i\xi ;  \label{Fcompo} \\
&&  \notag \\
&(\Lambda ,\Sigma )=&(0,i)\;\;:\mathcal{F}_{0i}(X)=-\frac{1}{2}d_{ijk}\frac{%
X^{j}X^{k}}{(X^{0})^{2}}; \\
&&  \notag \\
&(\Lambda ,\Sigma )=&(i,j)\;\;:\mathcal{F}_{ij}(X)=d_{ijk}\frac{X^{k}}{X^{0}}%
,
\end{eqnarray}
or in matrix form:
\begin{equation}
\mathcal{F}_{\Lambda \Sigma }=\left(
\begin{array}{ccc}
F_{00}(X;\xi ) &  & F_{0j}(X) \\
&  &  \\
F_{i0}(X) &  & F_{ij}(X)
\end{array}
\right) =\left(
\begin{array}{ccc}
\frac{1}{3}d_{ijk}\frac{X^{i}X^{j}X^{k}}{(X^{0})^{3}}+2i\xi &  & -\frac{1}{2}%
d_{jkl}\frac{X^{k}X^{l}}{(X^{0})^{2}} \\
&  &  \\
-\frac{1}{2}d_{ikl}\frac{X^{k}X^{l}}{(X^{0})^{2}} &  & d_{ijk}\frac{X^{k}}{%
X^{0}}
\end{array}
\right) .  \label{fmatrix}
\end{equation}
Notice that only $\mathcal{F}_{00}$ changes by additive constant ``$2i\xi $%
'' with respect to the classical case of $d$-SK geometry ($\xi =0$).

In order to explicitly compute the various terms of Eq. (\ref{N-matrix}),
let us start observing that Eq. (\ref{fmatrix}) yields
\begin{equation}
Im\left[ \mathcal{F}_{\Lambda \Sigma }(X;\xi )\right] =\left(
\begin{array}{ccc}
\frac{1}{3}d_{ijk}Im\left[ \frac{X^{i}X^{j}X^{k}}{(X^{0})^{3}}\right] +2\xi
&  & -\frac{1}{2}d_{jkl}Im\left[ \frac{X^{k}X^{l}}{(X^{0})^{2}}\right] \\
&  &  \\
-\frac{1}{2}d_{ikl}Im\left[ \frac{X^{k}X^{l}}{(X^{0})^{2}}\right] &  &
d_{ijk}Im\left[ \frac{X^{k}}{X^{0}}\right]
\end{array}
\right) .  \label{imf}
\end{equation}
Through Eqs. (\ref{fmatrix}) and (\ref{imf}), one can then compute:
\begin{eqnarray}
Im\left[ \mathcal{F}_{\Lambda \Sigma }(X;\xi )\right] X^{\Lambda }X^{\Sigma
} &=&\left( \frac{1}{3}d_{ijk}Im\left[ \frac{X^{i}X^{j}X^{k}}{(X^{0})^{3}}%
\right] +2\xi \right) (X^{0})^{2}+  \notag \\
&&-d_{ikl}Im\left[ \frac{X^{k}X^{l}}{(X^{0})^{2}}\right] X^{0}X^{i}+d_{ijk}Im%
\left[ \frac{X^{k}}{X^{0}}\right] X^{i}X^{j};  \notag \\
&& \\
\left( Im\mathcal{F}_{\Lambda \Omega }\right) \left( Im\mathcal{F}_{\Sigma
\Psi }\right) X^{\Omega }X^{\Psi } &=&\left( Im\mathcal{F}_{\Lambda
0}\right) \left( Im\mathcal{F}_{\Sigma 0}\right) (X^{0})^{2}+\left( Im%
\mathcal{F}_{\Lambda 0}\right) \left( Im\mathcal{F}_{\Sigma j}\right)
X^{0}X^{j}+  \notag \\
&&+\left( Im\mathcal{F}_{\Lambda i}\right) \left( Im\mathcal{F}_{\Sigma
0}\right) X^{0}X^{i}+\left( Im\mathcal{F}_{\Lambda i}\right) \left( Im%
\mathcal{F}_{\Sigma j}\right) X^{i}X^{j}.  \notag \\
&&
\end{eqnarray}

Within the assumptions (\ref{special-coords})-(\ref{Kgf}), one can thus
write:
\begin{equation}
\mathcal{F}_{\Lambda \Sigma }(z;\xi )=\left(
\begin{array}{ccc}
\frac{1}{3}d_{ijk}z^{i}z^{j}z^{k}+2i\xi &  & -\frac{1}{2}d_{jkl}z^{k}z^{l}
\\
&  &  \\
-\frac{1}{2}d_{ikl}z^{k}z^{l} &  & d_{ijk}z^{k}
\end{array}
\right) ,
\end{equation}
yielding Eq. (\ref{imfsymplectic}). Through the definition (\ref{z-split-def}%
), one can further elaborate as follows:
\begin{eqnarray}
d_{ijk}Im\left( z^{k}\right) &=&-d_{ijk}\lambda ^{k};  \label{s-1} \\
d_{ikl}Im\left( z^{k}z^{l}\right) &=&-2d_{ikl}x^{k}\lambda ^{l};  \label{s-2}
\\
d_{ijk}Im\left( z^{i}z^{j}z^{k}\right) &=&-d_{ijk}\left( 3x^{i}x^{j}\lambda
^{k}-\lambda ^{i}\lambda ^{j}\lambda ^{k}\right) .  \label{s-3}
\end{eqnarray}
Thus, the denominator of the second term in Eq. (\ref{N-matrix}) explicitly
reads:
\begin{equation}
Im\left[ \mathcal{F}_{\Lambda \Sigma }(X;\xi )\right] X^{\Lambda }X^{\Sigma
}=\frac{4}{3}d_{ijk}\lambda ^{i}\lambda ^{j}\lambda ^{k}+2\xi .
\end{equation}

On the other hand, the $(\Lambda ,\Sigma =0,0)$-component of the numerator
of the second term in Eq. (\ref{N-matrix}) reads
\begin{eqnarray}
Im\left( \mathcal{F}_{0\Omega }\right) Im\left( \mathcal{F}_{0\Delta
}\right) X^{\Omega }X^{\Delta } &=&\left( \frac{1}{3}d_{ijk}\lambda
^{i}\lambda ^{j}\lambda ^{k}-id_{ijk}x^{i}\lambda ^{j}\lambda ^{k}\right)
^{2}+  \notag \\
&&+4\xi ^{2}+\frac{4}{3}\xi d_{ijk}\lambda ^{i}\lambda ^{j}\lambda
^{k}-4i\xi d_{ijk}x^{i}\lambda ^{j}\lambda ^{k}.
\end{eqnarray}
Thus, through Eqs. (\ref{N-matrix}), (\ref{imfsymplectic}) and (\ref{s-1})-(%
\ref{s-3}), the following result is achieved:
\begin{eqnarray}
\mathcal{N}_{00} &=&\overline{\mathcal{F}}_{00}+2i\frac{\left( Im\mathcal{F}%
_{0\Omega }\right) \left( Im\mathcal{F}_{0\Delta }\right) X^{\Omega
}X^{\Delta }}{Im\left[ \mathcal{F}_{\Theta \Xi }(X;\lambda )\right]
X^{\Theta }X^{\Xi }}=  \notag \\
&=&\frac{1}{3}d_{ijk}x^{i}x^{j}x^{k}+9\xi \frac{d_{ijk}x^{i}\lambda
^{j}\lambda ^{k}}{\left( 2d_{pqr}\lambda ^{p}\lambda ^{q}\lambda ^{r}+3\xi
\right) }+  \notag \\
&&+i\left\{ d_{ijk}x^{i}x^{j}\lambda ^{k}-\frac{1}{6}d_{ijk}\lambda
^{i}\lambda ^{j}\lambda ^{k}-\frac{1}{4}\xi +3\frac{\left[ \frac{9}{4}\xi
^{2}-(d_{ijk}x^{i}\lambda ^{j}\lambda ^{k})^{2}\right] }{\left(
2d_{pqr}\lambda ^{p}\lambda ^{q}\lambda ^{r}+3\xi \right) }\right\} .
\label{N00}
\end{eqnarray}

It is worth noticing that in the classical limit $\xi \rightarrow 0$ one
reobtains the known result for $\mathcal{N}_{00}$ in $d$-SKG \cite{CFM1}
(see also App. A of \cite{Cardoso-N}).

Similarly, the $(\Lambda ,\Sigma =0,i)$-component of the numerator of the
second term in Eq. (\ref{N-matrix}) can be computed to be:
\begin{equation}
\left( Im\mathcal{F}_{0\Omega }\right) \left( Im\mathcal{F}_{i\Delta
}\right) X^{\Omega }X^{\Delta }=d_{ijk}\lambda ^{j}\lambda ^{k}\left(
d_{pqr}x^{p}\lambda ^{q}\lambda ^{r}+\frac{i}{3}d_{pqr}\lambda ^{p}\lambda
^{q}\lambda ^{r}+2i\xi \right) .
\end{equation}
Thus, through Eqs. (\ref{N-matrix}), (\ref{imfsymplectic}) and (\ref{s-1})-(%
\ref{s-3}), one obtains:
\begin{eqnarray}
\mathcal{N}_{0i} &=&\overline{\mathcal{F}}_{0i}+2i\frac{\left( Im\mathcal{F}%
_{0\Omega }\right) \left( Im\mathcal{F}_{i\Delta }\right) X^{\Omega
}X^{\Delta }}{Im\left[ \mathcal{F}_{\Theta \Xi }(X;\lambda )\right]
X^{\Theta }X^{\Xi }}=  \notag \\
&=&-\frac{1}{2}d_{ijk}x^{j}x^{k}-\frac{3}{2}\frac{\xi }{(\frac{2}{3}%
d_{lmn}\lambda ^{l}\lambda ^{m}\lambda ^{n}+\xi )}d_{ijk}\lambda ^{j}\lambda
^{k}+  \notag \\
&&+i\left[ -d_{ijk}x^{j}\lambda ^{k}+\frac{d_{pqr}x^{p}\lambda ^{q}\lambda
^{r}}{(\frac{2}{3}d_{lmn}\lambda ^{l}\lambda ^{m}\lambda ^{n}+\xi )}%
d_{ijk}\lambda ^{j}\lambda ^{k}\right] ,  \label{N0i}
\end{eqnarray}
which in the classical limit $\xi \rightarrow 0$ is consistent with the
known result for $\mathcal{N}_{0i}$ in $d$-SKG \cite{CFM1} (see also App. A
of \cite{Cardoso-N}).

The expression of $\mathcal{N}_{ij}$ is (almost) the same of the classical ($%
\lambda \rightarrow 0$) case \cite{CFM1} (see also App. A of \cite{Cardoso-N}%
), namely:
\begin{eqnarray}
\mathcal{N}_{ij} &=&\overline{\mathcal{F}}_{ij}+2i\frac{\left( Im\mathcal{F}%
_{i\Omega }\right) \left( Im\mathcal{F}_{j\Delta }\right) X^{\Omega
}X^{\Delta }}{Im\left[ \mathcal{F}_{\Theta \Xi }(X;\lambda )\right]
X^{\Theta }X^{\Xi }}=  \notag \\
&=&d_{ijk}x^{k}+i\left[ d_{ijk}\lambda ^{k}-\frac{d_{ikm}d_{jln}\lambda
^{k}\lambda ^{m}\lambda ^{l}\lambda ^{n}}{(\frac{2}{3}d_{pqr}\lambda
^{p}\lambda ^{q}\lambda ^{r}+\xi )}\right] .  \label{Nij}
\end{eqnarray}

In order to compute $V_{BH}$ by using Eq. (\ref{VBH2}) and subsequent
definitions, it is convenient to introduce the symmetric matrix
\begin{equation}
\mathcal{A}_{ij}\equiv \frac{1}{12}\frac{\nu ^{1/3}}{\widetilde{\nu }}\left(
\widehat{d}_{ij}-\frac{\nu }{\widetilde{\nu }}\frac{\widehat{d}_{i}\widehat{d%
}_{j}}{4}\right) ,  \label{A-matrix}
\end{equation}
whose inverse reads
\begin{equation}
\mathcal{A}^{ij}=12\frac{\widetilde{\nu }}{\nu ^{1/3}}\left[ \widehat{d}%
^{ij}-\frac{\nu ^{2/3}\widehat{\lambda }^{i}\widehat{\lambda }^{j}}{2\left(
\nu -\frac{\xi }{2}\right) }\right] ,~~\mathcal{A}^{ij}\mathcal{A}%
_{jk}\equiv \delta _{k}^{i},
\end{equation}
along with the related contractions
\begin{eqnarray}
\mathcal{A}_{i} &\equiv &\mathcal{A}_{ij}x^{j}; \\
\mathcal{A} &\equiv &\mathcal{A}_{ij}x^{i}x^{j}.
\end{eqnarray}

Notice that there is no simple relations between the symmetric matrices $%
g_{ij}$ and $\mathcal{A}_{ij}$, respectively given by Eqs. (\ref{covmetric})
and (\ref{A-matrix}). Generally, they are proportional only in the classical
limit:
\begin{eqnarray}
\lim_{\xi \rightarrow 0}\mathcal{A}_{ij} &=&\frac{1}{12}\nu ^{-2/3}\left(
\widehat{d}_{ij}-\frac{\widehat{d}_{i}\widehat{d}_{j}}{4}\right) =-\frac{1}{3%
}\breve{g}_{ij};  \label{1} \\
\lim_{\xi \rightarrow 0}\mathcal{A}^{ij} &=&12\nu ^{2/3}\left( \widehat{d}%
^{ij}-\frac{\widehat{\lambda }^{i}\widehat{\lambda }^{j}}{2}\right) =-3%
\breve{g}^{ij}.  \label{2}
\end{eqnarray}

Finally, one can then compute the following expressions:
\begin{eqnarray}
Re\mathcal{N}_{00} &=&\frac{1}{3}h+\frac{3}{4}\xi \frac{\nu ^{2/3}}{%
\widetilde{\nu }}\widehat{d}_{i}x^{i};  \label{R-00} \\
Re\mathcal{N}_{0i} &=&-\frac{1}{2}h_{i}-\frac{3}{8}\xi \frac{\nu ^{2/3}}{%
\widetilde{\nu }}\widehat{d}_{i}=-\frac{1}{2}\frac{\partial Re\mathcal{N}%
_{00}}{\partial x^{i}};  \label{R-0i} \\
Re\mathcal{N}_{ij} &=&h_{ij}=\frac{1}{2}\frac{\partial ^{2}Re\mathcal{N}_{00}%
}{\partial x^{i}\partial x^{j}};  \label{R-ij} \\
&&  \notag \\
Im\mathcal{N}_{00} &=&-\widetilde{\nu }\left( 1-12\mathcal{A}\right) +\frac{9%
}{16}\frac{\xi ^{2}}{\widetilde{\nu }};  \label{I-00} \\
Im\mathcal{N}_{0i} &=&-12\widetilde{\nu }\mathcal{A}_{i}=-\frac{1}{2}\frac{%
\partial Im\mathcal{N}_{00}}{\partial x^{i}};  \label{I-0i} \\
Im\mathcal{N}_{ij} &=&12\widetilde{\nu }\mathcal{A}_{ij}=\frac{1}{2}\frac{%
\partial ^{2}Im\mathcal{N}_{00}}{\partial x^{i}\partial x^{j}};  \label{I-ij}
\\
&&  \notag \\
\left( Im\mathcal{N}\right) ^{-1\mid 00} &=&-\frac{1}{\widetilde{\nu }}%
\left( 1-\frac{9}{16}\frac{\xi ^{2}}{\widetilde{\nu }^{2}}\right) ^{-1};
\label{I-1-00} \\
\left( Im\mathcal{N}\right) ^{-1\mid 0i} &=&-\frac{1}{\widetilde{\nu }}%
\left( 1-\frac{9}{16}\frac{\xi ^{2}}{\widetilde{\nu }^{2}}\right) ^{-1}x^{i};
\label{I-1-0i} \\
\left( Im\mathcal{N}\right) ^{-1\mid ij} &=&-\frac{1}{\widetilde{\nu }}%
\left( 1-\frac{9}{16}\frac{\xi ^{2}}{\widetilde{\nu }^{2}}\right) ^{-1}\left[
x^{i}x^{j}-\frac{1}{12}\left( 1-\frac{9}{16}\frac{\xi ^{2}}{\widetilde{\nu }%
^{2}}\right) \mathcal{A}^{ij}\right] .  \label{I-1-ij}
\end{eqnarray}
The equivalent matrix expressions of Eqs. (\ref{R-00})-(\ref{I-1-ij})
respectively read
\begin{eqnarray}
Re\mathcal{N}_{\Lambda \Sigma } &=&\left(
\begin{array}{ccc}
\frac{1}{3}h+\frac{3}{4}\xi \frac{\nu ^{2/3}}{\tilde{\nu}}\widehat{d}%
_{i}x^{i} &  & -\frac{1}{2}h_{j}-\frac{3}{8}\xi \frac{\nu ^{2/3}}{\tilde{\nu}%
}\widehat{d}_{j} \\
&  &  \\
-\frac{1}{2}h_{i}-\frac{3}{8}\xi \frac{\nu ^{2/3}}{\tilde{\nu}}\widehat{d}%
_{i} &  & h_{ij}
\end{array}
\right) ;  \label{R-matrix} \\
&&  \notag \\
Im\mathcal{N}_{\Lambda \Sigma } &=&-\tilde{\nu}\left(
\begin{array}{ccc}
1-12\mathcal{A}-\frac{9}{16}\frac{\xi ^{2}}{\tilde{\nu}^{2}} &  & 12\mathcal{%
A}_{j} \\
&  &  \\
12\mathcal{A}_{i} &  & -12\mathcal{A}_{ij}
\end{array}
\right) ;  \label{I-matrix} \\
&&  \notag \\
\left( Im\mathcal{N}_{\Lambda \Sigma }\right) ^{-1} &=&-\frac{1}{\tilde{\nu}}%
\left( 1-\frac{9}{16}\frac{\xi ^{2}}{\tilde{\nu}^{2}}\right) ^{-1}\left(
\begin{array}{ccc}
1 &  & x^{j} \\
&  &  \\
x^{i} &  & x^{i}x^{j}-\frac{1}{12}\left( 1-\frac{9}{16}\frac{\xi ^{2}}{%
\tilde{\nu}^{2}}\right) \mathcal{A}^{ij}
\end{array}
\right) .  \notag \\
&&  \label{I-1-matrix}
\end{eqnarray}
In the classical limit ($\xi \rightarrow 0$) all above expressions yield the
known results for $d$-SKG \cite{CFM1} (see also App. A of \cite{Cardoso-N}).

All above results then yield to the explicit expression of $V_{BH}$ given by
Eq. (\ref{V_BH-d-SKG-lambda}).

\section{Details of Computation of $E$-Tensor}

In order to compute the $E$-tensor for the prepotential (\ref{effe}), let us
start by splitting the differential operator $\partial $ according to Eq. (%
\ref{z-split-def}):
\begin{equation}
\partial _{i}\equiv \frac{\partial }{\partial z^{i}}=\frac{1}{2}\left( \frac{%
\partial }{\partial x^{i}}+i\frac{\partial }{\partial \lambda ^{i}}\right) .
\label{d-split-def}
\end{equation}
This, by its very definition (\ref{C}), the $C$-tensor can be computed to be
(see also Eq. (\ref{K}))
\begin{equation}
C_{ijk}=\exp \left( K\right) d_{ijk}=\frac{3}{4}\frac{d_{ijk}}{\left(
d_{lmn}\lambda ^{l}\lambda ^{m}\lambda ^{n}-3\xi \right) }=\frac{1}{8}\frac{%
d_{ijk}}{\left( \nu -\frac{\xi }{2}\right) },
\end{equation}
thus implying that the $C$-tensor simply gets multiplicatively renormalized
with respect to its classical ($\xi \rightarrow 0$) expression:
\begin{equation}
C_{ijk}=\frac{\nu }{\left( \nu -\frac{\xi }{2}\right) }C_{ijk,\xi =0}.
\end{equation}
Thence, by observing that

\begin{eqnarray}
\partial _{i}d_{jk} &=&\frac{i}{2}\frac{\partial }{\partial \lambda ^{i}}%
d_{jk}=\frac{i}{2}d_{ijk};  \label{ss-1} \\
\partial _{i}d_{j} &=&\frac{i}{2}\frac{\partial }{\partial \lambda ^{i}}%
d_{j}=2d_{ij}=2\nu ^{1/3}\widehat{d}_{ij};  \label{ss-2} \\
\partial _{i}\nu &=&\frac{i}{2}\frac{\partial }{\partial \lambda ^{i}}\nu =%
\frac{i}{4}d_{i}=\frac{i}{4}\nu ^{2/3}\widehat{d}_{i},  \label{ss-3}
\end{eqnarray}
one can compute that
\begin{equation}
\partial _{i}C_{jkl}=-\frac{i}{2^{5}}\frac{\nu }{\left( \nu -\frac{\xi }{2}%
\right) ^{2}}\widehat{d}_{i}\widehat{d}_{jkl},  \label{partialC}
\end{equation}
and (consistent with Eq. (\ref{K})):
\begin{equation}
\partial _{i}K=-\frac{i}{4}\frac{\nu ^{2/3}}{\left( \nu -\frac{\xi }{2}%
\right) }\widehat{d}_{i}.  \label{partialK}
\end{equation}
Using Eqs. (\ref{covmetric}), (\ref{contrmetric}) and (\ref{Gamma}), the
connection $\Gamma $ is computed as follows:
\begin{eqnarray}
-i\frac{2}{3}\Gamma _{ij}^{~~m} &=&-\frac{\nu ^{2/3}}{6(\nu +\xi )}\widehat{d%
}_{ij}\widehat{\lambda }^{m}-\frac{5}{24}\frac{\nu ^{8/3}}{(\nu +\xi )\left(
\nu -\frac{\xi }{2}\right) ^{2}}\widehat{d}_{i}\widehat{d}_{j}\widehat{%
\lambda }^{m}+  \notag \\
&&+\frac{\nu ^{2/3}}{6\left( \nu -\frac{\xi }{2}\right) }\widehat{d}%
_{i}\delta _{j}^{m}+\frac{\nu ^{2/3}}{6\left( \nu -\frac{\xi }{2}\right) }%
\widehat{d}_{j}\delta _{i}^{m}-\frac{1}{3}d_{ijl}\widehat{d}^{ml}\nu ^{-1/3}
\notag \\
&=&i\frac{2}{3}\Gamma _{\left( ij\right) }^{~~m},
\end{eqnarray}
and therefore:
\begin{eqnarray}
-i\frac{16}{3}\left( \nu -\frac{\xi }{2}\right) \Gamma _{ij}^{~~m}C_{mkl}
&=&-\frac{\nu ^{2/3}}{6\left( \nu +\xi \right) }\widehat{d}_{ij}\widehat{d}%
_{kl}-\frac{5}{24}\frac{\nu ^{5/3}}{(\nu +\xi )\left( \nu -\frac{\xi }{2}%
\right) ^{2}}\widehat{d}_{i}\widehat{d}_{j}\widehat{d}_{kl}+  \notag \\
&&+\frac{\nu ^{2/3}}{6\left( \nu -\frac{\xi }{2}\right) }\widehat{d}%
_{i}d_{jkl}+\frac{\nu ^{2/3}}{6\left( \nu -\frac{\xi }{2}\right) }\widehat{d}%
_{j}d_{ikl}-\frac{1}{3}\nu ^{-1/3}d_{ijn}d_{mkl}\widehat{d}^{mn}.  \notag \\
&&
\end{eqnarray}
Through some straightforward elaborations, all this leads to Eq. (\ref
{covderC}), and then to Eqs. (\ref{E-csi-1}) and (\ref{ETfinal}).

\section{Details of Computation of Riemann Tensor}

In order to compute the Riemann tensor $R_{i\overline{j}k\overline{l}}$
given by Eq. (\ref{Riemann-Kahler}) (\textit{i.e.} working in the approach
considered in Subsect. \ref{First-Approach}), one needs to recall Eqs. (\ref
{K}), (\ref{covmetric}), (\ref{contrmetric}), (\ref{d-split-def}) and (\ref
{ss-1})-(\ref{ss-3}), which leads to the following results:
\begin{eqnarray}
\overline{\partial }_{\overline{l}}\overline{\partial }_{\overline{j}%
}\partial _{m}K &=&\overline{\partial }_{\overline{l}}g_{m\overline{j}}=-%
\frac{i}{2}\frac{\partial }{\partial \lambda ^{l}}g_{m\overline{j}}=  \notag
\\
&=&\frac{i}{8}\frac{1}{\left( \nu -\frac{\xi }{2}\right) }\left[ d_{mjl}-%
\frac{\nu }{2\left( \nu -\frac{\xi }{2}\right) }\left( \widehat{d}_{mj}%
\widehat{d}_{l}+\widehat{d}_{ml}\widehat{d}_{j}+\widehat{d}_{jl}\widehat{d}%
_{m}\right) +\frac{\nu ^{2}}{4\left( \nu -\frac{\xi }{2}\right) ^{2}}%
\widehat{d}_{m}\widehat{d}_{j}\widehat{d}_{l}\right] ;  \notag \\
&& \\
\overline{\partial }_{\overline{l}}\partial _{i}\overline{\partial }_{%
\overline{j}}\partial _{k}K &=&\overline{\partial }_{\overline{l}}\partial
_{i}g_{k\overline{j}}=-\frac{i}{2}\frac{\partial }{\partial \lambda ^{l}}%
\left( \partial _{i}g_{k\overline{j}}\right) =  \notag \\
&=&\frac{1}{32}\frac{\nu ^{2/3}}{\left( \nu -\frac{\xi }{2}\right) ^{2}}%
\left[
\begin{array}{l}
\left( d_{ijk}\widehat{d}_{l}+d_{ijl}\widehat{d}_{k}+d_{ikl}\widehat{d}%
_{j}+d_{jkl}\widehat{d}_{i}\right) + \\
\\
+2\left( \widehat{d}_{ij}\widehat{d}_{kl}+\widehat{d}_{ik}\widehat{d}_{jl}+%
\widehat{d}_{il}\widehat{d}_{jk}\right) + \\
\\
-\frac{\nu }{\left( \nu -\frac{\xi }{2}\right) }\left(
\begin{array}{l}
\widehat{d}_{ij}\widehat{d}_{k}\widehat{d}_{l}+\widehat{d}_{ik}\widehat{d}%
_{j}\widehat{d}_{l}+\widehat{d}_{jk}\widehat{d}_{i}\widehat{d}_{l}+ \\
+\widehat{d}_{il}\widehat{d}_{j}\widehat{d}_{k}+\widehat{d}_{jl}\widehat{d}%
_{i}\widehat{d}_{k}+\widehat{d}_{kl}\widehat{d}_{i}\widehat{d}_{j}
\end{array}
\right) + \\
\\
+\frac{3}{4}\frac{\nu ^{2}}{\left( \nu -\frac{\xi }{2}\right) ^{2}}\widehat{d%
}_{i}\widehat{d}_{j}\widehat{d}_{k}\widehat{d}_{l}
\end{array}
\right] ;  \notag \\
&&
\end{eqnarray}
\begin{eqnarray}
g^{m\overline{n}}\left( \overline{\partial }_{\overline{l}}\overline{%
\partial }_{\overline{j}}\partial _{m}K\right) \partial _{i}\overline{%
\partial }_{\overline{n}}\partial _{k}K &=&g^{m\overline{n}}\left( \overline{%
\partial }_{\overline{l}}g_{m\overline{j}}\right) \partial _{i}g_{k\overline{%
n}}=-\frac{1}{96}\frac{\nu ^{2/3}}{\left( \nu -\frac{\xi }{2}\right) ^{2}}%
\cdot  \notag \\
&&\cdot \left[
\begin{array}{l}
-3\left( d_{ijk}\widehat{d}_{l}+d_{ijl}\widehat{d}_{k}+d_{ikl}\widehat{d}%
_{j}+d_{jkl}\widehat{d}_{i}\right) + \\
-9\frac{\left( \nu +\frac{\xi }{2}\right) }{\nu +\xi }\widehat{d}_{ik}%
\widehat{d}_{jl}+ \\
+\frac{3}{2}\frac{\left( 2\nu ^{2}+\nu \xi -\xi ^{2}\right) \nu }{\left( \nu
-\frac{\xi }{2}\right) ^{2}\left( \nu +\xi \right) }\left( \widehat{d}_{i}%
\widehat{d}_{k}\widehat{d}_{jl}+\widehat{d}_{j}\widehat{d}_{l}\widehat{d}%
_{ik}\right) + \\
+\frac{3}{2}\frac{\nu }{\left( \nu -\frac{\xi }{2}\right) }\left( \widehat{d}%
_{k}\widehat{d}_{l}\widehat{d}_{ij}+\widehat{d}_{i}\widehat{d}_{l}\widehat{d}%
_{jk}+\widehat{d}_{j}\widehat{d}_{k}\widehat{d}_{il}+\widehat{d}_{i}\widehat{%
d}_{j}\widehat{d}_{kl}\right) + \\
-\frac{1}{4}\frac{\left( 2\nu ^{2}+\nu \xi -\xi ^{2}\right) \nu ^{2}}{\left(
\nu -\frac{\xi }{2}\right) ^{3}\left( \nu +\xi \right) }\widehat{d}_{i}%
\widehat{d}_{j}\widehat{d}_{k}\widehat{d}_{l}+ \\
+6\frac{\left( \nu -\frac{\xi }{2}\right) }{\nu }d_{ijk}d_{jlm}\widehat{d}%
^{mn}
\end{array}
\right] .  \notag \\
&&
\end{eqnarray}
All above results are the basic ingredients, through some simple algebra, of
the result (\ref{Riemann-csi}).


\begin{thebibliography}{99}
\bibitem{FKS}  S. Ferrara, R. Kallosh and A. Strominger, $\mathcal{N}\mathit{%
=2}$\textit{\ Extremal Black Holes}, Phys. Rev. \textbf{D52}, 5412 (1995),
\texttt{hep-th/9508072}.

\bibitem{Strom}  A. Strominger, \textit{Macroscopic Entropy of }$\mathcal{N}%
\mathit{=2}$\textit{\ Extremal Black Holes}, Phys. Lett. \textbf{B383}, 39
(1996), \texttt{hep-th/9602111}.

\bibitem{FK1}  S. Ferrara and R. Kallosh, \textit{\ Supersymmetry and
Attractors}, Phys. Rev. \textbf{D54}, 1514 (1996), \texttt{hep-th/9602136}.

\bibitem{FK2}  S. Ferrara and R. Kallosh, \textit{\ Universality of
Supersymmetric Attractors}, Phys. Rev. \textbf{D54}, 1525 (1996), \texttt{%
hep-th/9603090}.

\bibitem{FGK}  S.~Ferrara, G. W. Gibbons and R. Kallosh, \textit{Black Holes
and Critical Points in Moduli Space}, Nucl. Phys. \textbf{B500}, 75 (1997),
\texttt{hep-th/9702103}.

\bibitem{BPS}  G. W. Gibbons and C. M. Hull, \textit{A Bogomol'ny Bound for
General Relativity and Solitons in }$\mathcal{N}\mathit{=2}$\textit{\
Supergravity}, Phys. Lett. \textbf{B109}, 190 (1982).

\bibitem{ADFT-review}  L. Andrianopoli, R. D'Auria, S. Ferrara and M.
Trigiante, \textit{Extremal black holes in supergravity}, Lect. Notes Phys.
\textbf{737}, 661 (2008), \texttt{hep-th/0611345}.

\bibitem{Sen-review}  A. Sen, \textit{Black Hole Entropy Function,
Attractors and Precision Counting of Microstates}, \texttt{arXiv:0708.1270}.

\bibitem{Kallosh-review}  S. Bellucci, S. Ferrara, R. Kallosh and A.
Marrani, \textit{Extremal Black Hole and Flux Vacua Attractors}, Lect. Notes
Phys. \textbf{755}, 115 (2008), \texttt{arXiv:0711.4547 [hep-th]}.

\bibitem{Erice-07}  S. Ferrara, K. Hayakawa and A. Marrani, \textit{Lectures
on Attractors and Black Holes}, Fortsch. Phys. \textbf{56}, 993 (2008),
\texttt{arXiv:0805.2498 [hep-th]}.

\bibitem{BFGM2}  S. Bellucci, S. Ferrara, M. G\"{u}naydin and A. Marrani,
\textit{SAM Lectures on Extremal Black Holes in }$\mathit{d=4}$\textit{\
Extended Supergravity}, \texttt{arXiv:0905.3739 [hep-th]}.

\bibitem{Sen:2005wa}  A.~Sen, \textit{Black hole entropy function and the
attractor mechanism in higher derivative gravity}, JHEP \textbf{0509}, 038
(2005), \texttt{hep-th/0506177}.

\bibitem{Strominger-SKG}  A. Strominger, \textit{Special Geometry}, Commun.
Math. Phys. \textbf{133}, 163 (1990).

\bibitem{CDF-review}  A. Ceresole, R. D'Auria and S. Ferrara, \textit{The
Symplectic Structure of }$\mathcal{N}\mathit{=2}$ \textit{Supergravity and
Its Central Extension}, Nucl. Phys. Proc. Suppl. \textbf{46} (1996), \texttt{%
hep-th/9509160}.

\bibitem{Freed}  D. S. Freed, \textit{Special K\"{a}hler manifolds}, Commun.
Math. Phys. \textbf{203}, 31 (1999), \texttt{hep-th/9712042}.

\bibitem{BFGM1}  S. Bellucci, S. Ferrara, M. G\"{u}naydin and A. Marrani,
\textit{Charge orbits of symmetric special geometries and attractors}, Int.
J. Mod. Phys. \textbf{A21}, 5043 (2006), \texttt{hep-th/0606209}.

\bibitem{Mohaupt-1}  G. L. Cardoso, D. Lust and T. Mohaupt, \textit{Modular
symmetries of }$\mathcal{N}\mathit{=2}$\textit{\ black holes}, Phys. Lett.
\textbf{B388}, 266 (1996), \texttt{hep-th/9608099}.

\bibitem{CQ-N=2-BHs}  K. Behrndt, G. L. Cardoso, B. de Wit, R. Kallosh, D.
Lust and T. Mohaupt, \textit{Classical and quantum }$\mathcal{N}\mathit{=2}$%
\textit{\ supersymmetric black holes}, Nucl. Phys. \textbf{B488}, 236
(1997), \texttt{hep-th/9610105}.

\bibitem{Behrndt-1}  K. Behrndt, \textit{Quantum corrections for }$\mathit{%
D=4}$\textit{\ black holes and }$\mathit{D=5}$\textit{\ strings}, Phys.
Lett. \textbf{B396}, 77 (1997), \texttt{hep-th/9610232}.

\bibitem{Behrndt-2}  K. Behrndt and T. Mohaupt, \textit{Entropy of }$%
\mathcal{N}\mathit{=2}$\textit{\ black holes and their }$\mathit{M}$\textit{%
-brane description}, Phys. Rev. \textbf{D56}, 2206 (1997), \texttt{%
hep-th/9611140}.

\bibitem{Behrndt-3}  K. Behrndt, G. L. Cardoso, I. Gaida, \textit{Quantum }$%
\mathcal{N}\mathit{=2}$\textit{\ supersymmetric black holes in the }$\mathit{%
S-T}$\textit{\ model}, Nucl. Phys. \textbf{B506}, 267 (1997), \texttt{%
hep-th/9704095}.

\bibitem{Maldacena}  J.~M.~Maldacena, A.~Strominger and E.~Witten, \textit{%
Black hole entropy in M-theory}, JHEP \textbf{9712}, 002 (1997), \texttt{%
hep-th/9711053}.

\bibitem{Mohaupt-2}  G. L. Cardoso, B. de Wit and T. Mohaupt, \textit{%
Corrections to macroscopic supersymmetric black hole entropy}, Phys. Lett.
\textbf{B451}, 309 (1999), \texttt{hep-th/9812082}.

\bibitem{Cardoso-0}  G.~L. Cardoso, B.~de Wit and T.~Mohaupt, \textit{%
Deviations from the area law for supersymmetric black holes,} Fortsch.\
Phys. \textbf{48}, 49 (2000), \texttt{hep-th/9904005}.

\bibitem{Mohaupt-3}  G. L. Cardoso, B. de Wit and T. Mohaupt, \textit{%
Macroscopic entropy formulae and nonholomorphic corrections for
supersymmetric black holes}, Nucl. Phys. \textbf{B567}, 87 (2000), \texttt{%
hep-th/9906094}.

\bibitem{R-square-1}  G. L. Cardoso, B. de Wit, J. Kappeli and T. Mohaupt,
\textit{Stationary BPS solutions in }$\mathcal{N}\mathit{=2}$\textit{\
supergravity with }$\mathit{R}^{2}$\textit{\ interactions}, JHEP \textbf{0012%
}, 019 (2000), \texttt{hep-th/0009234}.

\bibitem{Lust-1}  G. L. Cardoso, D. L\"{u}st and J. Perz, \textit{Entropy
maximization in the presence of higher-curvature interactions}, JHEP \textbf{%
0605}, 028 (2006), \texttt{hep-th/0603211}.

\bibitem{Lust-2}  G. L. Cardoso, V. Grass, D. L\"{u}st and J. Perz, \textit{%
Extremal non-BPS Black Holes and Entropy Extremization}, JHEP \textbf{0609},
078 (2006), \texttt{hep-th/0607202}.

\bibitem{Cardoso-1}  G.L. Cardoso, B. de Wit and S. Mahapatra, \textit{Black
hole entropy functions and attractor equations}, JHEP \textbf{0703}, 085
(2007), \texttt{hep-th/0612225}.

\bibitem{Cardoso-2}  G.L. Cardoso, B. de Wit and S. Mahapatra, \textit{%
Subleading and non-holomorphic corrections to }$\mathcal{N}\mathit{=2}$%
\textit{\ BPS black hole entropy}, HEP \textbf{0902}, 006 (2009), \texttt{%
arXiv:0808.2627 [hep-th]}.

\bibitem{CFG}  S. Cecotti, S. Ferrara and L. Girardello, \textit{Geometry of
Type }$\mathit{II}$\textit{\ Superstrings and the Moduli of Superconformal
Field Theories}, Int. J. Mod. Phys. \textbf{A4}, 2475 (1989).

\bibitem{Peccei-Quinn}  R. D. Peccei and H. R. Quinn, \textit{Constraints
imposed by CP conservation in the presence of instantons}, Phys. Rev.
\textbf{D16}, 1791 (1977). R. D. Peccei and H. R. Quinn, \textit{CP
conservation in the presence of instantons}, Phys. Rev. Lett. \textbf{38},
1440 (1977). R. D. Peccei and H. R. Quinn, \textit{Some aspects of instantons%
}, Nuovo Cim. \textbf{A41}, 309 (1977).

\bibitem{dWVP}  B. de Wit and A. Van Proeyen, \textit{Special geometry,
cubic polynomials and homogeneous quaternionic spaces}, Commun. Math. Phys.
\textbf{149}, 307 (1992), \texttt{hep-th/9112027}.

\bibitem{dWVVP}  B. de Wit, F. Vanderseypen and A. Van Proeyen, \textit{%
Symmetry structure of special geometries}, Nucl. Phys. \textbf{B400}, 463
(1993), \texttt{hep-th/9210068}..

\bibitem{CDLOGP1}  P. Candelas, X. C. De La Ossa, P. S. Green and L. Parkes,
\textit{A Pair of Calabi-Yau Manifolds as an Exactly Soluble Superconformal
Theory}, Nucl. Phys. \textbf{B359}, 21 (1991). P. Candelas, X. C. De La
Ossa, P. S. Green and L. Parkes, \textit{An Exactly Soluble Superconformal
Theory from a Mirror Pair of Calabi-Yau Manifolds}, Phys. Lett. \textbf{B258}%
, 118 (1991).

\bibitem{Alvarez-Gaume}  L. Alvarez-Gaume, D. Z. Freedman, \textit{%
Geometrical Structure and Ultraviolet Finiteness in the Supersymmetric Sigma
Model}, Commun. Math. Phys. \textbf{80}, 443 (1981).

\bibitem{Grisaru}  M. T. Grisaru, A. van de Ven and D. Zanon, \textit{Four
Loop Beta Function for the }$\mathcal{N}\mathit{=1}$\textit{\ and }$\mathcal{%
N}\mathit{=2}$\textit{\ Supersymmetric Nonlinear Sigma Model in Two
Dimensions}, Phys. Lett. \textbf{B173}, 423 (1986). M. T. Grisaru, A. van de
Ven and D. Zanon, \textit{Two Dimensional Supersymmetric Sigma Models on
Ricci Flat K\"{a}hler Manifolds are not Finite}, Nucl. Phys. \textbf{B277},
388 (1986). M. T. Grisaru, A. van de Ven and D. Zanon, \textit{Four Loop
Divergences for the }$\mathcal{N}\mathit{=1}$\textit{\ Supersymmetric
Nonlinear Sigma Model in Two Dimensions}, Nucl. Phys. \textbf{B277}, 409
(1986).

\bibitem{HKTY}  S. Hosono, A. Klemm, S. Theisen and Shing-Tung Yau, \textit{%
Mirror symmetry, mirror map and applications to Calabi-Yau hypersurfaces},
Commun. Math. Phys. \textbf{167}, 301 (1995), \texttt{hep-th/9308122}.

\bibitem{Witten-theta}  E. Witten, \textit{Dyons of charge e theta/2 pi},
Phys. Lett. \textbf{B86}, 283 (1979).

\bibitem{FHSV}  S. Ferrara, J. A. Harvey, A. Strominger and C. Vafa, \textit{%
Second quantized mirror symmetry}, Phys. Lett. \textbf{B361}, 59 (1995),
\texttt{hep-th/9505162}. P. S. Aspinwall, \textit{An }$\mathcal{N}\mathit{=2}
$\textit{\ dual pair and a phase transition}, Nucl. Phys. \textbf{B460}, 57
(1996), \texttt{hep-th/9510142}. D. R. Morrison and C. Vafa, \textit{%
Compactifications of F theory on Calabi-Yau threefolds. 1}, Nucl. Phys.
\textbf{B473}, 74 (1996), \texttt{hep-th/9602114}. D. R. Morrison and C.
Vafa, \textit{Compactifications of F theory on Calabi-Yau threefolds. 2},
Nucl. Phys. \textbf{B476}, 437 (1996), \texttt{hep-th/9603161}. J. A. Harvey
and G. W. Moore, \textit{Exact gravitational threshold correction in the
FHSV model}, Phys. Rev. \textbf{D57}, 2329 (1998), \texttt{hep-th/9611176}.
A. Klemm and M. Marino, \textit{Counting BPS states on the Enriques
Calabi-Yau}, Commun. Math. Phys. \textbf{280}, 27 (2008), \texttt{%
hep-th/0512227}. J. R. David, \textit{On the dyon partition function in }$%
\mathcal{N}\mathit{=2}$\textit{\ theories}, JHEP \textbf{0802}, 025 (2008),
\texttt{arXiv:0711.1971}.

\bibitem{Ferrara-Bianchi}  M. Bianchi and S. Ferrara, \textit{Enriques and
Octonionic Magic Supergravity Models}, JHEP \textbf{0802}, 054 (2008),
\texttt{arXiv:0712.2976 [hep-th]}.

\bibitem{Gunaydin-lectures}  M. G\"{u}naydin, \textit{Lectures on Spectrum
Generating Symmetries and }$\mathit{U}$\textit{-duality in Supergravity,
Extremal Black Holes, Quantum Attractors and Harmonic Superspace}, \texttt{%
arXiv:0908.0374 [hep-th]}.

\bibitem{BFMY}  S. Bellucci, S. Ferrara, A. Marrani and A. Yeranyan, \textit{%
Mirror Fermat Calabi-Yau Threefolds and Landau-Ginzburg Black Hole Attractors%
}, Riv. Nuovo Cim. \textbf{029}, 1 (2006), \texttt{hep-th/0608091}.

\bibitem{Misra1}  P. Kaura and A. Misra, \textit{On the Existence of
Non-Supersymmetric Black Hole Attractors for Two-Parameter Calabi-Yau's and
Attractor Equations}, Fortsch. Phys. \textbf{54}, 1109 (2006), \texttt{%
hep-th/0607132}.

\bibitem{BFMS1}  S. Bellucci, S. Ferrara, A. Marrani and A. Shcherbakov,
\textit{Splitting of Attractors in }$1$\textit{-modulus Quantum Corrected
Special Geometry}, JHEP \textbf{0802}, 088 (2008), \texttt{arXiv:0710.3559
[hep-th]}.

\bibitem{Kal1}  A. Chou, R. Kallosh, J. Rahmfeld, Soo-Jong Rey, M. Shmakova
and Wing Kai Wong, \textit{Critical points and phase transitions in 5-d
compactifications of M-theory}, Nucl. Phys. \textbf{B508}, 147 (1997),
\texttt{hep-th/9704142}.

\bibitem{Kal2}  R. Kallosh, A. D. Linde and M. Shmakova, \textit{%
Supersymmetric multiple basin attractors}, JHEP \textbf{9911}, 010 (1999),
\texttt{hep-th/9910021}.

\bibitem{Zhukov}  M. Wijnholt and S. Zhukov, \textit{On the Uniqueness of
Black Hole Attractors}, \texttt{hep-th/9912002}.

\bibitem{Moore}  G.W. Moore, \textit{Attractors and Arithmetic}, \texttt{%
hep-th/9807056}. G.W. Moore, \textit{Arithmetic and Attractors}, \texttt{%
hep-th/9807087}. G.W. Moore, \textit{Les Houches Lectures on Strings and
Arithmetic}, \texttt{hep-th/0401049}.

\bibitem{Ferrara-Marrani-1}  S. Ferrara and A. Marrani, $\mathcal{N}\mathit{%
=8}$\textit{\ non-BPS Attractors, Fixed Scalars and Magic Supergravities},
Nucl. Phys.~\textbf{B788}, 63 (2008), \texttt{arXiV:0705.3866}.

\bibitem{ferrara4}  S. Ferrara and A. Marrani, \textit{On the Moduli Space
of non-BPS Attractors for~}$\mathcal{N}\mathit{=2}$\textit{\ Symmetric
Manifolds}, Phys. Lett.~\textbf{B652}, 111 (2007), \texttt{arXiV:0706.1667}.

\bibitem{BFSY-1}  S. Bellucci, S. Ferrara, A. Shcherbakov and A. Yeranyan,
\textit{Black hole entropy, flat directions and higher derivatives}, \texttt{%
arXiv:0906.4910 [hep-th]}.

\bibitem{CVP}  E. Cremmer and A. Van Proeyen, \textit{Classification Of
Kahler Manifolds In }$\mathcal{N}\mathit{=2}$\textit{\ Vector Multiplet
Supergravity Couplings}, Class. Quant. Grav. \textbf{2}, 445 (1985).

\bibitem{BFMS2}  S. Bellucci, S. Ferrara, A. Marrani and A. Shcherbakov,
\textit{Quantum Lift of Non-BPS Flat Directions}, Phys. Lett. \textbf{B672},
77 (2009), \texttt{arXiv:0811.3494 [hep-th]}.

\bibitem{Duff-stu}  M. J. Duff, J. T. Liu and J. Rahmfeld, \textit{%
Four-dimensional string/string/string triality}, Nucl. Phys. \textbf{B459},
125 (1996), \texttt{hep-th/9508094}.

\bibitem{BKRSW}  K. Behrndt, R. Kallosh, J. Rahmfeld, M. Shmakova and W. K.
Wong, \textit{STU Black Holes and String Triality}, Phys. Rev. \textbf{D54},
6293 (1996), \texttt{hep-th/9608059}.

\bibitem{Shmakova}  M. Shmakova, \textit{Calabi-Yau black holes}, Phys. Rev.
\textbf{D56}, 540 (1997), \texttt{hep-th/9612076}.

\bibitem{TT1}  P. K. Tripathy and S. P. Trivedi, \textit{Non-supersymmetric
attractors in string theory}, JHEP \textbf{0603}, 022 (2006), \texttt{%
hep-th/0511117}.

\bibitem{Saraikin-Vafa-1}  K. Saraikin and C. Vafa, \textit{%
Non-supersymmetric Black Holes and Topological Strings}, Class. Quant. Grav.
\textbf{25}, 095007 (2008), \texttt{hep-th/0703214}.

\bibitem{TT2}  S. Nampuri, P. K. Tripathy and S. P. Trivedi, \textit{On The
Stability of Non-Supersymmetric Attractors in String Theory}, JHEP \textbf{%
0708}, 054 (2007), \texttt{arXiV:0705.4554}.

\bibitem{BMOS-1}  S. Bellucci, A. Marrani, E. Orazi and A. Shcherbakov,
\textit{Attractors with Vanishing Central Charge}, Phys. Lett. \textbf{B655}%
, 185 (2007), \texttt{arXiV:0707.2730}.

\bibitem{stu-unveiled}  S. Bellucci, S. Ferrara, A. Marrani and A. Yeranyan,
$\mathit{stu}$\textit{\ Black Holes Unveiled}, Entropy Vol. \textbf{10}(4),
507 (2008), \texttt{arXiv:0807.3503 [hep-th]}.

\bibitem{AFMT1}  L. Andrianopoli, S. Ferrara, A. Marrani and M. Trigiante,
\textit{Non-BPS Attractors in }$\mathit{5d}$\textit{\ and }$\mathit{6d}$%
\textit{\ Extended Supergravity}, Nucl. Phys. \textbf{B795}, 428 (2008),
\texttt{arXiv:0709.3488}.

\bibitem{CFM1}  A.Ceresole, S.Ferrara and A. Marrani, $\mathit{4d/5d}$%
\textit{\ Correspondence for the Black Hole Potential and its Critical
Points,} Class. Quant. Grav. \textbf{24}, 5651 (2007), \texttt{%
arXiV:0707.0964 [hep-th]}.

\bibitem{FerBig}  L.~Andrianopoli, M.~Bertolini, A.~Ceresole, R.~D'Auria,
S.~Ferrara, P.~Fr\'{e} and T.~Magri, $\mathcal{N}\mathit{=2}$\textit{\
supergravity and }$\mathcal{N}\mathit{=2}$\textit{\ super Yang-Mills theory
on general scalar manifolds: Symplectic covariance, gaugings and the
momentum map,} J.\ Geom.\ Phys.\ \textbf{23}, 111 (1997), \texttt{%
hep-th/9605032}. L. Andrianopoli, M. Bertolini, A. Ceresole, R. D'Auria, S.
Ferrara and P. Fr\'{e}, \textit{General Matter Coupled }$\mathcal{N}\mathit{%
=2}$ \textit{Supergravity}, Nucl. Phys. \textbf{B476}, 397 (1996), \texttt{%
hep-th/9603004}.

\bibitem{Castellani1}  L. Castellani, R. D'Auria and S. Ferrara, \textit{%
Special Geometry without Special Coordinates}, Class. Quant. Grav. \textbf{7}%
, 1767 (1990). L. Castellani, R. D'Auria and S. Ferrara, \textit{Special
K\"{a}hler Geometry: an Intrinsic Formulation from }$\mathcal{N}\mathit{=2}$%
\textit{\ Space-Time Supersymmetry}, Phys. Lett. \textbf{B241}, 57 (1990).

\bibitem{DFF}  R. D'Auria, S. Ferrara and P. Fr\'{e}, \textit{Special and
Quaternionic Isometries: General Couplings in }$\mathcal{N}\mathit{=2}$%
\textit{\ Supergravity and the Scalar Potential}, Nucl. Phys. \textbf{B359},
705 (1991).

\bibitem{Valencia}  S. Bellucci, S. Ferrara and A. Marrani, \textit{%
Attractors in Black}, Fortsch. Phys. \textbf{56}, 761 (2008), \texttt{%
ArXiv:0805.1310} \texttt{[hep-th]}.

\bibitem{Lu}  Z. Lu, \textit{A Note on Special K\"{a}hler Manifolds}, Math.
Ann. \textbf{313}, 711 (1999), \texttt{math/0505577}.

\bibitem{CFMZ1}  B. L. Cerchiai, S. Ferrara, A. Marrani, and B. Zumino,
\textit{Duality, Entropy and ADM Mass in Supergravity}, Phys. Rev. \textbf{%
D79}, 125010 (2009), \texttt{arXiv:0902.3973 [hep-th]}.

\bibitem{DFT-hom-non-symm}  R. D'Auria, S. Ferrara and M. Trigiante, \textit{%
Critical points of the Black-Hole potential for homogeneous special
geometries}, JHEP \textbf{0703}, 097 (2007), \texttt{hep-th/0701090}.

\bibitem{GRLS-1}  M. Gomez-Reino, J. Louis and C. A. Scrucca, \textit{No
metastable de Sitter vacua in }$\mathcal{N}=2$\textit{\ supergravity with
only hypermultiplets}, JHEP \textbf{0902}, 003 (2009), \texttt{%
arXiv:0812.0884}.

\bibitem{BH1}  J. D. Bekenstein, Phys. Rev. \textbf{D7}, 2333 (1973). S. W.
Hawking, Phys. Rev. Lett. \textbf{26}, 1344 (1971); in C. DeWitt, B. S.
DeWitt, \textit{Black Holes (Les Houches 1972)} (Gordon and Breach, New
York, 1973). S. W. Hawking, Nature \textbf{248}, 30 (1974). S. W. Hawking,
Comm. Math. Phys. \textbf{43}, 199 (1975).

\bibitem{Helgason}  S. Helgason, \textit{Differential Geometry, Lie Groups
and Symmetric Spaces} (Academic Press, New York, 1978).

\bibitem{LA08-Proc}  S. Ferrara and A. Marrani, \textit{Symmetric Spaces in
Supergravity}, contribution to the Proceedings of \textit{``Symmetry in
Mathematics and Physics''}, Los Angeles, CA, Jan 18-20, 2008, \texttt{%
arXiv:0808.3567 [hep-th]}.

\bibitem{FK-N=8}  S. Ferrara and R. Kallosh, \textit{On }$\mathcal{N}\mathit{%
=8}$\textit{\ Attractors}, Phys. Rev. \textbf{D73}, 125005 (2006), \texttt{%
hep-th/0603247}.

\bibitem{ADF-Duality-d=4}  L. Andrianopoli, R. D'Auria and S. Ferrara, $%
\mathit{U}$\textit{\ invariants, black hole entropy and fixed scalars},
Phys. Lett. \textbf{B403}, 12 (1997), \texttt{hep-th/9703156}.

\bibitem{Bern}  Z. Bern, J. J. Carrasco, L. J. Dixon, H. Johansson, D. A.
Kosower and R. Roiban, \textit{Three-Loop Finiteness of }$\mathcal{N}\mathit{%
=8}$\textit{\ Supergravity}, Phys. Rev. Lett. \textbf{98}, 161303 (2007),
\texttt{hep-th/0702112}.

\bibitem{K-N=8}  R. Kallosh, \textit{On UV Finiteness of the Four Loop }$%
\mathcal{N}\mathit{=8}$\textit{\ Supergravity}, JHEP \textbf{0909}, 116
(2009), \texttt{ArXiv:0906.3495 [hep-th]}.

\bibitem{Cardoso-N}  G. L. Cardoso, J. M. Oberreuter and J. Perz, \textit{%
Entropy function for rotating extremal black holes in very special geometry}%
, JHEP \textbf{0705}, 025 (2007), \texttt{hep-th/0701176}.
\end{thebibliography}
\end{document}